\documentclass[longauth]{aa}
\makeatletter
\renewcommand*\aa@pageof{, page \thepage{} of \pageref*{LastPage}}
\makeatother
\usepackage{txfonts}
\usepackage{graphicx}
\usepackage{ifthen}
\usepackage{mathptmx}
\usepackage[breaklinks=true]{hyperref}
\usepackage{xcolor}
\usepackage{color, colortbl}
\usepackage{amsmath}
\usepackage{xspace}
\usepackage{placeins}
\usepackage{natbib}
\usepackage{longtable}
\usepackage{tablefootnote}
\usepackage{array}
\hypersetup{colorlinks,linkcolor={blue},citecolor={blue},urlcolor={blue}}
\newcommand{\fermi}{{\em Fermi}\xspace}
\newcommand{\swift}{{\em Swift}\xspace}

\newcommand{\EP}{{\em EP}\xspace}
\newcommand{\svom}{{\em SVOM}\xspace}
\newcommand{\grb}{GRB~140304A\xspace}
\newcommand{\Ep}{$E_{\rm p}$\xspace}
\newcommand{\tninty}{{T$_{90}$}\xspace}
\newcommand{\tzero}{{T$_{0}$}\xspace}
\newcommand{\sw}[1][default]{\texttt{#1}}
\bibpunct{(}{)}{;}{a}{}{,}

\begin{document}
\title{High-redshift GRB 140304A at \(z\) = 5.282 with flaring activity: A multi-wavelength study}

\titlerunning{\grb: Multi-Wavelength flare at \(z=5.282\)}
\author{Shashi. B. Pandey\inst{1} \thanks{E-mail: shashiaries0@gmail.com}
\and Amit. K. Ror\inst{1} \thanks{E-mail: mitturor77894@gmail.com}
\and A. J. Castro-Tirado\inst{3,10}
\and A. Pozanenko\inst{8,9}
\and V. Lipunov\inst{6,7}
\and S. Jeong\inst{2}
\and I. H. Park\inst{11}
\and R. S\'anchez-Ramírez\inst{3,12}
\and B.-B. Zhang\inst{13}
\and D. Xu\inst{14}
\and N. R. Butler\inst{15}
\and C. G. Mundell\inst{16}
\and S. R. Oates\inst{17}
\and P. Y. Minaev\inst{8}
\and A. Volnova\inst{8}
\and J. Bai\inst{18}
\and J. Bloom\inst{19}
\and N. Budnev\inst{20}
\and A. Castell\'on\inst{21}
\and Ch. Cui\inst{14}
\and M. D. Caballero-García\inst{3}
\and Maria Gritsevich\inst{32,33}
\and G. Garc\'ia-Segura\inst{26}
\and D. Hiriart\inst{26}
\and A. Valeev\inst{34}
\and S. Castillo-Carrion\inst{22}
\and Yash Sharma\inst{1, 35}
\and Y. Fan\inst{18}
\and E. Gorbovskoy\inst{6}
\and O. Gress\inst{20}
\and S. Guziy\inst{3}
\and Y.-D. Hu\inst{3}
\and Brajesh Kumar\inst{36,37}
\and Amar Aryan\inst{38}
\and Rahul Gupta\inst{39}
\and E. V. Klunko\inst{24}
\and V. Kornilov\inst{25,7}
\and A. Kutyrev\inst{4}
\and G. Antipov\inst{6}
\and A. Kuznetsov\inst{6}
\and William H. Lee\inst{26}
\and C. P\'erez del Pulgar\inst{10}
\and R. Querel\inst{27}
\and M. G. Richer\inst{26}
\and S. E. Schmalz\inst{28}
\and N. Tiurina\inst{6}
\and N. Tungalag\inst{29}
\and K. Zhirkov\inst{6}
\and A. M. Watson\inst{30}
\and Ch. Wang\inst{18,31}
\and P. Balanutsa\inst{6}}

\institute{Aryabhatta Research Institute of Observational Sciences (ARIES), Manora Peak, Nainital-263002, India
\and Department of Physics, Sungkyunkwan University (SKKU), 2066 Seobu-ro, Suwon, 440-746, Republic of Korea
\and Instituto de Astrofísica de Andalucía (IAA-CSIC), Glorieta de la Astronomía s/n, E-18008, Granada, Spain
\and Astrophysics Science Division, NASA Goddard Space Flight Center, Greenbelt, MD 20771, USA
\and Korea Astronomy and Space Science Institute, Daejeon 34055, Republic of Korea
\and Lomonosov Moscow State University, SAI, 119234 Moscow, Universitetsky pr. 13, Russia
\and Lomonosov Moscow State University, Physics Department, 119991, Moscow, Vorobievy hills, 1, Russia
\and Space Research Institute of the Russian Academy of Sciences, 117997, Profsoyuzanaya, 84/32 Moscow, Russia
\and National Research University Higher School of Economics, Myasnitskaya 20, 101000, Moscow, Russia
\and Departamento de Ingeniería de Sistemas y Automática, Escuela de Ingenierías, Universidad de Málaga, Dr. Pedro Ortiz Ramos, 29071 Málaga, Spain
\and Department of Physics, Sungkyunkwan University (SKKU), 2066 Seobu-ro, Suwon, 440-746, Republic of Korea
\and INAF, Istituto Astrofísica de Planetologia Spaziali, Via Fosso del Cavaliere 100, I-00133 Roma, Italy
\and School of Astronomy and Space Science, Nanjing University, Nanjing 210093, China
\and CAS Key Laboratory of Space Astronomy and Technology, National Astronomical Observatories, Chinese Academy of Sciences, Beijing 100101, China
\and School of Earth and Space Exploration, Arizona State University, Tempe, AZ 85287, USA
\and European Space Astronomy Centre, Villafranca del Castillo, 28691, Villanueva de la Canada, Madrid, Spain
\and Physics Department, Lancaster University, Bailrigg, Lancaster LA1 4YB, UK
\and Yunan Astronomical Observatories, CAS Kunming 650011, Yunnan, China
\and Astronomy Department, University of California, Berkeley, CA 94720-7450, USA
\and Irkutsk State University, Applied Physics Institute, 20, Gagarin Blvd, 664003, Irkutsk, Russia
\and Facultad de Ciencias, Universidad de Málaga, Bulevard Louis Pasteur, 29010 Málaga, Spain
\and Servicio Central de Informática, Universidad de Málaga, Bulevard Louis Pasteur, 29010 Málaga, Spain
\and Instituto de Astronomía, Universidad Nacional Autónoma de México, Apartado Postal 70-264, 04510 México, D. F., México
\and Institute of Solar-Terrestrial Physics, 664033, p/o box 291, Lermontov st., 126a, Irkutsk, Russia
\and Lomonosov Moscow State University, SAI, 119234 Moscow, Universitetsky pr. 13, Russia
\and Instituto de Astronomía, Unidad Académica Ensenada, Universidad Autónoma de México, Ensenada BC 22760, México
\and Earth Sciences, 1 Fairway Drive, Avalon, Lower Hutt 5011 New Zealand
\and Keldysh Institute of Applied Mathematics, Moscow, Russia
\and Institute of Astronomy and Geophysics, Mongolian Academy of Sciences, Ulaanbaatar, Mongolia
\and Instituto de Astronomía, Universidad Nacional Autónoma de México, Apartado Postal 70-264, 04510 México, D. F., México
\and University of Chinese Academy of Sciences, Beijing, 100049, China
\and Faculty of Science, University of Helsinki, Gustaf Hallströmin katu 2, FI-00014 Helsinki, Finland
\and Institute of Physics and Technology, Ural Federal University, Mira str. 19, 620002 Ekaterinburg
\and Special Astrophysical Observatory of Russian Academy of Sciences, Nizhniy Arkhyz 369167, RussiaSpecial Astrophysical Observatory, Zelenchuck, Russia
\and Indian Institute of Technology, IIT-Roorkee, Uttarakhand, India, 247667
\and South-Western Institute for Astronomy Research, Yunnan University, Kunming, Yunnan 650500, People's Republic of China
\and Yunnan Key Laboratory of Survey Science, Yunnan University, Kunming, Yunnan 650500, People's Republic of China
\and Graduate Institute of Astronomy, National Central University, 300 Jhongda Road, 32001 Jhongli, Taiwan
\and Astrophysics Science Division, NASA Goddard Space Flight Center, Mail Code 661, Greenbelt, MD 20771, USA}

\date{Received XXX; accepted XXXX}

\newpage

\abstract
{This article presents a detailed multi-wavelength analysis of \grb at $z=5.282$, having uncommon late-time flaring features. The aim is to study \grb and other similar bursts to understand stellar evolution and formation processes at high-$z$.}
{GRBs at high-$z$, possible flaring activities at different frequencies seen at relatively late-times, help to constrain temporal correlation among contemporaneous flares. In the present study, we plan to constrain such a temporal and spectral study for a sample of high-$z$ bursts, including \grb.}
{We use \swift, \fermi, and ground-based observations to constrain the temporal and spectral properties of the prompt and afterglow emissions. Using the cross-correlation function, we calculate the spectral lag in the light curves observed in two energy bands of \swift's Burst Alert Telescope (BAT) and X-ray Telescope (XRT).}
{Parameter evolution of the prompt emission analysis reveals a hard-to-soft evolution of the spectral peak energy (\Ep) and the magnetic field strength (B), consistent with the typical population of long GRBs. For \grb, a rare pattern of spectral lag evolution having positive lag in the early BAT light curves, but no lag is observed in the XRT light curves. We have also observed systematic time delays among the peak times of flares in three different bands, but the optical flares exhibit a morphological correspondence with X/$\gamma$-ray flares.}
{Our analysis shows that the observed positive spectral lag in \grb is closely related to the hard-to-soft spectral evolution during the prompt emission phase, as seen in some of the other long GRBs. Additionally, there is a clear connection between $\gamma$/X-ray and optical flares with prompt emission, which are produced through synchrotron radiation during rapid bulk acceleration within the emitting region.}


\maketitle
\section{Introduction} \label{sec:introduction}
Gamma-ray bursts (GRBs) are among the most luminous explosions, enabling their detection at cosmological distances. Theoretical predictions for the range of redshift over which GRBs can be detected within the capabilities of current telescopes are \(0<z<20\) \citep{2000ApJ...536....1L, 2024A&A...686A..56K}, with the furthest detected GRB lying at $z \sim 9.4$ \citep{2011ApJ...736....7C, 2018ApJ...865..107T, Rossi_2022, 2025A&A...695A.239B}. GRB emission occurs in two successive phases, the prompt emission, during which highly variable $\gamma$-ray or X-ray emission dominates within the duration (\tninty \footnote{Time interval over which 90\% of the fluence is detected in $\gamma$/X-rays}) ranges from \(10^{-3}-10^{4}\)\,s. Following this, a broadband (radio-optical-X/$\gamma$-rays) afterglow emission is observed with a duration of a few hours to a few months \citep{1994ApJ...430L..93R, 1997ApJ...490...92K, 2015PhR...561....1K}. The afterglow emission occurs when the fireball interacts with the surrounding medium, producing external shocks: the forward shock (moving outward) and the reverse shock (travelling backwards), which heat the ejecta within the fireball. According to the predictions of typical shock models, afterglow light curve and spectra are assumed to decay following largely power-law relations of various kinds under diverse circumstances both temporally and spectrally \citep{Sari_1998, 2002ApJ...571..779P, 2006ApJ...637..889Z, 2009MNRAS.395..490O, 2012MNRAS.426L..86O, 2013NewAR..57..141G, Racusin_2016, 2025MNRAS.543.2404R}.\\

Following the launch of the \textit{Neil Gehrels Swift Observatory} (\swift, \citealt{Gehrels_2004, 2006ApJ...639..303H}) in 2004, the complex behaviour of GRB's early X-ray afterglows was discovered, for example, a canonical X-ray light curve with {at least} five components: steep decay, plateau, normal decay, late jet break and flares \citep{2006ApJ...642..389N, 2006ApJ...647.1213O, 2006ApJ...642..354Z, 2005Sci...309.1833B, 2024BSRSL..93..709R}. The temporal and spectral properties of superimposed flares at different frequencies in many GRBs imply that these observed signatures may originate through a mechanism similar to the one proposed for the prompt emission and cannot be explained adequately by an external forward shock model \citep{2007ApJ...671.1921F, 2016ApJ...824L..16U}. Therefore, the early to late flares observed in the $\gamma$/X-ray/optical/near-infrared (NIR) light curves are indicative of the reactivation of the central engine offering a powerful diagnostic of emission mechanisms, shock physics and microphysical properties of the emitting plasma \citep{2008Natur.455..183R,2009ApJ...707.1623M, 2005ApJ...630L.113K, 2006MNRAS.370L..61P, 2010MNRAS.406.2149M, 2011MNRAS.417.2161B, 2013ApJ...774....2S, 2014ApJ...788...30S, 2021RAA....21..300Z, 2022MNRAS.513.2777K}. 

However, detecting such emission from high-redshift (i.e., \(z>5\)) GRBs is challenging due to their faintness. In other words, at such distances, their apparent brightness is lower and subject to the power-law decaying nature of GRB afterglows and their environments \citep{2024A&A...683A..55C, 2022NatAs...6.1101C}.
Nonetheless, investigating the origin of high-$z$ GRBs provides insight into stellar and galactic evolution and their environmental conditions in the early universe or over much of cosmic history \citep{2010MNRAS.401.2773S, 2015ApJ...804...51C, 2015JHEAp...7...35S, 2022ApJ...929..111F, 2025A&A...701A..84B}. Historically, Long GRBs (LGRBs, \tninty $>$ 2 s) are expected to originate from the collapse of massive Wolf-Rayet stars with their hydrogen and helium envelopes stripped off \citep{1993ApJ...405..273W}, except for a few recently detected nearby cases \citep{2021NatAs...5..917A, 2022Natur.612..228T, 2024Natur.626..737L}. In literature, the first generation of stars (Population-III $\rightleftharpoons$ Pop-III) could also be the possible progenitor of GRBs at high redshift \citep{2006MNRAS.369..825S, 2006ApJ...642..382B, 2026MNRAS.545f1985M}. Blue supergiant stars from Pop-III have also been proposed as potential progenitors to explain the prolonged central engine activity observed in Ultra-Long GRBs (ULGRBs, \tninty$>$1000\,s, \citealt{2018ApJ...859...48P, 2024ApJ...971..163R, 2025RMxAC..59..145A}).\\

To date, the \swift mission has been the most successful in detecting these high-$z$ bursts. \swift has detected a total of $\sim$1670 GRBs (according to the 3rd \swift-BAT catalogue; \citealt{2005SSRv..120..143B, 2016ApJ...829....7L}), but only about 518 (31\%) have measured redshifts, with an average redshift of LGRBs lies at about $z\sim2$ \citep{2019ApJS..245....1T, 2023Univ....9..113O}. Among these, $\sim$ 40 have photometric redshifts while the rest have spectroscopically measured redshifts obtained either from the burst itself or its host galaxy. From the complete set of \swift-GRBs, we extract a sub-sample of GRBs with $z>5$. In our sub-sample, there are $\sim$20 such high-$z$ GRBs, which are about 1\% of all \swift-detected bursts. Of these, 15 have spectroscopically confirmed redshifts, and 5 are constrained photometrically, illustrating the challenge of obtaining spectroscopic measurements due to their faint afterglow emission at such high redshifts. 

Building on the legacy of \swift, the recently launched \textit{Einstein Probe} (\EP; \citealt{2025SCPMA..6839501Y}) with its Wide-field X-ray Telescope (WXT) and the \textit{Space-based multi-band Variable Object Monitor} (\svom; \citealt{proceedingswei}) mission equipped with the ECLAIRs and Gamma Ray Monitor (GRM) instruments are dedicated to improving the detection of high-$z$ GRBs \citep{2014SPIE.9144E..24G, 2024A&A...685A.163L}. Thanks to their enhanced sensitivity extending to low X-ray energies and wide-field monitoring capabilities, these missions are expected to dramatically increase the discovery rate of GRBs from the early universe. Based on models calibrated with \swift data, these missions could detect around five GRBs per year at redshifts greater than 6 \citep{2025ApJ...988L..71W}. \svom’s strong potential is already demonstrated by the successful detection of GRB 250314A at redshift \(z=7.3\), corresponding to a time when the universe was only about 5\% of its current age \citep{2025A&A...704L...7C, 2026MNRAS.545f1985M}. Although \EP has not yet detected GRBs beyond redshift 5, it has observed bursts at redshift 4.8, showing promising capabilities \citep{2025NatAs...9..564L}. Together, \EP and \svom are dedicated to significantly expanding the sample of known high-$z$ GRBs. Their combined strengths will deepen our understanding of stellar evolution in the early universe and provide crucial insights into the reionisation era \citep{2025ApJ...988L..71W, 2025A&A...701A..84B, 2017RSOS....470304S, 2026ApJ...996..125W, 2025ApJS..281...20A}. The rarely observed subset of high-$z$ GRBs is an exceptional probe of dust properties in the early universe and the evolution of massive stars and the interstellar and intergalactic medium (\citealt{2005A&A...443..643Y, 2006A&A...460..199Y, 2007MNRAS.379.1589D, 2014A&A...564A..30G, 2024A&A...686L...4H} and references therein). The high-$z$ hosts are characterised as star-forming Lyman-break galaxies with minimal dust, and the majority of high-$z$ GRBs show low to moderate extinction with $\rm A_{V} < 1$ mag \citep{Zafar_sty1876, 2024A&A...690A.373R, 2018_Bolmer, 2024ApJ...966..133S, 2024A&A...691A.240M}.

In spite of decades of efforts mentioned above, detecting high-$z$ GRBs is still challenging, thus classifying them as a rather rare group of energetic cosmic events. \grb, one of these interesting bursts at $z = 5.282$, spectroscopically confirmed from the 10.4\,m Gran Telescopio Canarias (GTC) \citep{2014GCN.15936....1J}, is discussed as a part of the present analysis. Multi-wavelength observations, spanning the radio through $\gamma$-ray bands, and a rich, unpublished dataset from various ground-based telescopes (see the next section), make this burst particularly interesting for detailed analysis. In previous studies of multi-wavelength afterglow light curves of GRBs at high redshifts, the observed features (flares, plateaus etc) have been linked to several physical processes: the onset of forward and reverse shocks, episodes of energy injection, the presence of structured jets, and, in some cases originating from internal shock activity within the outflow \citep{2007ApJ...670..565L, 2008ApJ...675..528L, 2009MNRAS.395..490O, 2012ApJ...758...27L, 2013ApJ...774...13L, 2021MNRAS.505.4086G, 2023ApJ...942...34R}, a comprehensive discussion of the observed features can be found in \cite{2025MNRAS.543.2404R}. \grb at \(z=5.282\) is among the high-$z$ bursts having a rich multi-wavelength dataset, and the radio properties have already been investigated in \cite{2018ApJ...859..134L}, where bright radio flares were interpreted as signatures of reverse-shock emission and suggesting for a complex circumburst environment, with evidence for a dense, structured, \textit{Wind}-like medium. In this paper, we present new multi-wavelength observations of \grb, which show, for the first time, that multiple flares are observed over five orders in frequency from optical to $\gamma$-rays. \\

The article is structured as follows: \S \ref{sec:observations} describes the observations of \grb with a new data-set observed at diverse frequencies, \S \ref{sec:analysis} details the data reduction and presents the results, \S \ref{sec:discussion} discusses the findings. Finally, the summary and conclusions are given in \S \ref{sec:conclusion}. The cosmological parameters used in this paper are \( \rm H_0 = 71~km~s^{-1}~Mpc^{-1}\), \( \Omega_m = 0.30 \), and \( \Omega_\Lambda = 0.70 \) \citep{2001ApJ...553...47F}. 

\begin{figure}
\centering
\includegraphics[width=\columnwidth]{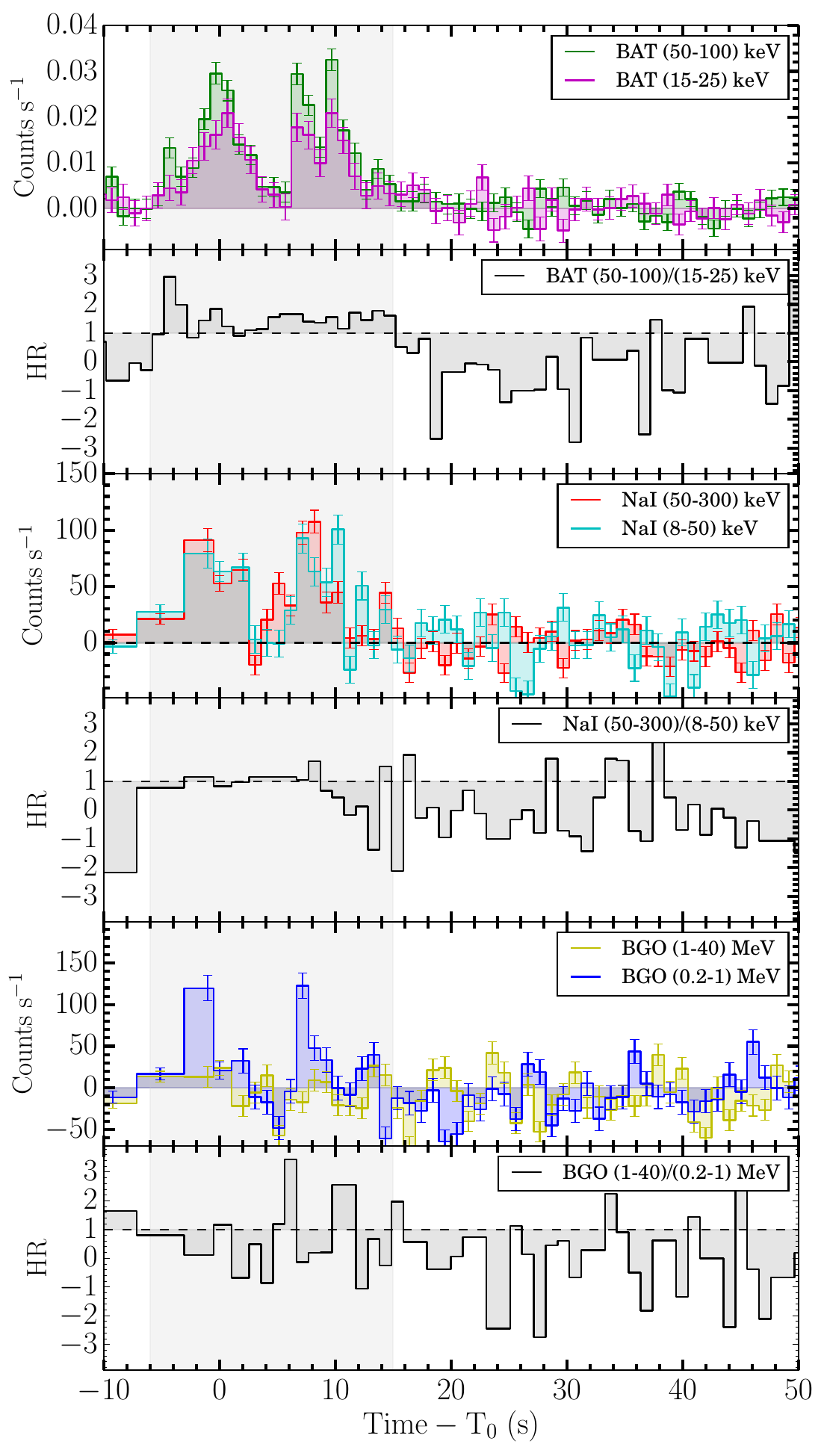}
\caption{The multi-band prompt emission background-subtracted light curve of \grb. The top two panels display the light curve observed by \swift-BAT and the corresponding hardness ratio. The middle two panels show the light curve from \fermi-GBM's NaI detector along with its associated hardness ratio. Similarly, the bottom two panels present the light curve from \fermi-GBM's BGO detector and the corresponding hardness ratio. The selected energy range from each instrument is shown in the legends.}
\label{fig:multiband_plc}
\end{figure}

\section{Observation details of \grb} \label{sec:observations}

\grb was first detected on 04 March 2014, by the \swift mission, and after this, observed by multiple space-based $\gamma$/X-ray instruments and ground-based optical, NIR, and radio telescopes, providing a comprehensive data set across the multi-wavelength regime.

\subsection{Space-based observations}
The details of space-based observations of \grb by \swift, \fermi, and others are given in the following subsections.

\subsubsection{\swift observations of \grb}
Burst Alert Telescope (BAT, \citealt{2005SSRv..120..143B}), the wide field $\gamma$-ray instrument onboard \swift, detected \grb at 13:22:31 UT (\tzero), pinpointing its coordinates to RA: 30.668$^\circ$, Dec: +33.485$^\circ$ (J2000), with an uncertainty radius of 3$'$ \citep{2014GCN.15915....1E}. BAT observations revealed the gamma-ray light curve has a multi-peaked structure with a \tninty duration of \(15.6 \pm 1.0\)\,s as measured in the 15-350\,keV, uncoded BAT energy range \citep{2014GCN.15914....1G, 2014GCN.15920....1M}. The time-averaged spectrum from \tzero-3.94 to \tzero+15.77\,s is well fit by a power-law function with a photon index of \(\Gamma = 1.29 \pm 0.08\) \citep{2014GCN.15927....1B}.\\

The {X-ray} Telescope (XRT, \citealt{2005SSRv..120..165B}) onboard \swift, commenced observations approximately \tzero+75\,s and detected a significant X-ray afterglow at the burst location, RA = 02h 02m 34.24s and Dec = +33$^{\circ}$ 28$^{'}$ 25.8$^{''}$ within errorbox of 1.6$^{''}$. The first 42\,s of \swift-XRT observations are conducted in window timing (WT) mode, and the rest of the observations exist in photon counting (PC) mode \citep{2014GCN.15925....1B, 2014GCN.15926....1P}.\\

Further, the Ultraviolet/Optical Telescope (UVOT, \citealt{2005SSRv..120...95R}) onboard \swift started settled observations 138\,s after the \swift-BAT trigger. Subsequent observations by \swift-UVOT involved multiple exposures with different filters, but no optical afterglow was detected in the observed field. Consequently, the instrument team provided only upper limits \citep{2014GCN.15920....1M}. Additionally, in our analysis, UVOT data were analysed in the \textit{v}-filter corresponding to the time of two bright flares, but no significant detections were found in any of the images.
 
\subsubsection{\fermi observations of \grb}
Gamma-ray Burst Monitor (GBM, \citealt{Meegan}) on board \fermi detected \grb at 13:22:31.48 UT, confirming the event independently. \fermi-GBM observation revealed a longer burst duration of \(32 \pm 6\)\,s \citep{2014GCN.15923....1J}. During the preliminary analysis of \fermi-GBM spectra, a \sw[Cutoff power-law] function is found to best describe the spectra with the low energy spectral index $\alpha$ = \(-1.1 \pm 0.22\) and spectral peak energy \Ep = \(185 \pm 35\)\,keV \citep{2014GCN.15923....1J}.

\subsection{Ground-based observations}
Several ground-based telescopes have observed the optical emission from \grb. At first, MASTER robotic telescopes initiated observations approximately 30\,s after the burst trigger. Next, Khureltogot, Nanshan, and Mondy observatories continued the observations. Reionisation and Transients Infrared Camera (RATIR) conducted observations and reported a \textit{g}-band dropout, suggesting its high-$z$ origin. The redshift of the burst at \(z = 5.282\) was constrained by the spectroscopic observations of 10.4\,m Gran Telescope Canarias (GTC) and Nordic Optical Telescope (NOT) \citep{2014GCN.15922....1J, 2014GCN.15921....1D, 2014GCN.15924....1D}. The complete details about the coordinated observations across different telescopes/missions and wavelengths can be found in the GCN Circulars for this GRB\footnote{\url{https://gcn.gsfc.nasa.gov/other/140304A.gcn3}}. 

\begin{figure}
\centering
\includegraphics[width=\columnwidth]{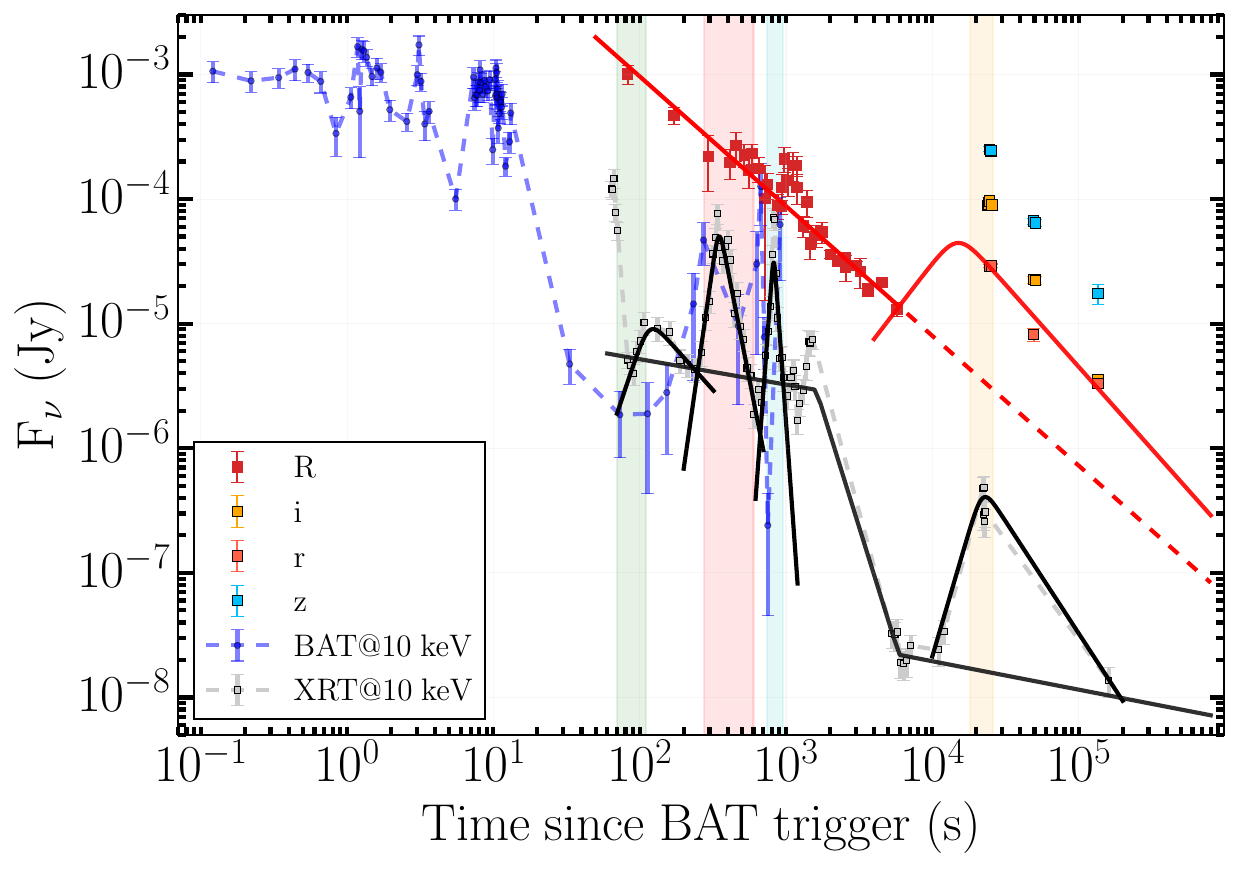}
\includegraphics[width=\columnwidth]{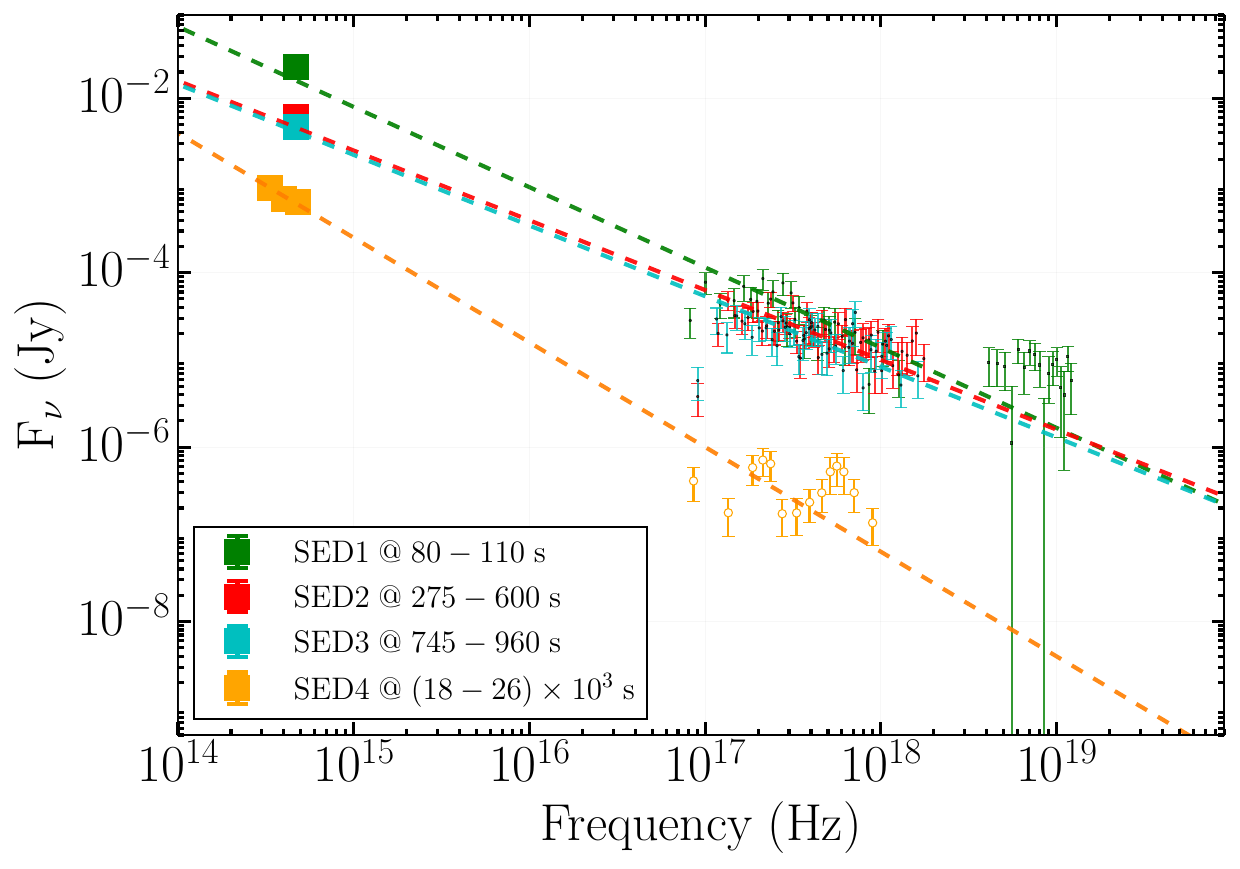}
\caption{Upper panel: the multi-band afterglow light curves of \grb. The \swift-BAT light curve is shown in blue, \swift-XRT in black, and the optical observations are shown with coloured squares, as shown in the legends. Lower panel: displays the multi-epoch optical to X-ray SEDs corresponding to the time intervals indicated by coloured strips in the upper panel. The optical observations are corrected both for Galactic and host extinctions, as described in section \ref{sec:extinction}.}
\label{fig:multiband_AGLC}
\end{figure}

\subsubsection{GTC spectroscopic observation}
Optical spectroscopy with OSIRIS at the 10.4\,m GTC started on 04 March 2014, starting $\sim$7.36\,h after the BAT trigger, using the R2500I VPHs (\(2 \times 1200\)\,s exposures). The 1.0$''$ slit was positioned on the location of the host galaxy, and 2 $\times$ 2 (pixels) binning mode was used for data acquisition. The obtained spectra were reduced and calibrated following standard procedures using custom tools based on the Image Reduction and Analysis Facility (IRAF; \citealt{1986SPIE..627..733T}). The obtained spectra had been flux-calibrated using observations of the spectro-photometric standard star, HILT600, observed on the same night with a 2.52$''$ slit. We did slit loss correction on the flux using an acquisition image taken with the same setup. The results from the spectral analysis are given in section \ref{sec:GTC_redshift}.

\subsubsection{MASTER optical observations}
The MASTER Global Robotic Net \citep{2010AdAst2010E..30L} responded to \grb 46\,s after the \swift-BAT trigger at MASTER-Amur observatory with 10\,s exposure obtained an upper limit of m$_{upperlim}$ $>$ 13.0 mag. The continued observation at MASTER-Tunka observatory started 84\,s after \swift-BAT trigger and detected an decaying optical afterglow, with index $\alpha_{opt}=0.77\pm0.05$, at burst location RA = 02h 02m 34.13s, Dec +33$^{\circ}$ 28$^{'}$ 26.6$^{''}$ \citep{2014GCN.15914....1G, 2014GCN.15932....1G}. All MASTER unfiltered magnitudes were calibrated as 0.2B + 0.8R by the USNO-B1 catalogue \citep{2012ExA....33..173K}. Subsequently, photometric data were processed with \sw[Astrokit] \citep{2014AstBu..69..368B} in order to minimise the standard deviations among the ensemble of comparison stars, and some stars with significant deviations were removed from the ensemble. Data for both polarisers were processed together, and the output data were cleared of instrumental effects, although some of the average common polarisation may remain for the reference stars. Tab. \ref{tab:optical_obs} provides the log of optical observations and photometry results. The MASTER data of the two polarisers are normalised to the Khureltogot data through comparisons of the detection at 1194.16\,s of MASTER and at 1180.90\,s of Khureltogot.

\subsubsection{Khureltogot and Mondy optical observations}
Observations of the optical afterglow of \grb with IKI-GRB-FuN \citep{Volnova_Alina} began on 04 March 2014, at 13:29:36 UT, i.e., $\sim$7 minutes later than the \swift-BAT trigger. The 40\,cm ORI-40 telescope of the Khureltogot observatory in Mongolia took 18 unfiltered images with exposures of 60\,s \citep{2014GCN.15918....1V}, and an afterglow of the magnitude of \(17.8 \pm 0.3\) mag was detected in the first frames. The observations continued at 13:54:10 UT, i.e., $\sim 0.5$ hours after the burst, with the 1.5-meter telescope AZT-33IK of Mondy observatory in Russia \citep{2014GCN.15917....1V}. We took 43 images in the \textit{R}-band with exposures of 60\,s, and the afterglow was clearly detected in stacked images. We continued observations with the AZT-33IK telescope on 05 March, starting at 13:17:34 UT. We took 60 frames with exposures of 120\,s using an \textit{I}-filter, but we did not detect the afterglow in a stacked image. Data were processed using the standard IRAF software package produced by the National Optical Astronomy Observatories (NOAO; \citealt{1986SPIE..627..733T}), CCDPROC (bias, dark reduction, flat field correction), IMAGES (sum and combining), and DAOPHOT (Aperture and point-spread function (PSF) photometry packages). The photometry is based on reference stars from the Sloan Digital Sky Survey (SDSS) Data Release 9 (DR9; \citealt{2012ApJS..203...21A}), where the \textit{R} mag, transformed from \textit{g,r,i}; Lupton 2005\footnote{\url{https://www.sdss3.org/dr10/algorithms/sdssUBVRITransform.php}}. The reference stars were selected using an automated online procedure for secondary photometric standards identification \citep{PankovNicolai2022}. The log of the observations and photometry results is listed in Tab. \ref{tab:optical_obs}. For cross-calibration of the Mondy observations, we added a goodness error by comparing the same reference star at the GTC Sloan-\textit{r} filter.

\subsubsection{Nanshan}
The 1\,m telescope located in Nanshan, Xinjiang, China began observations of \grb with the \textit{R}-filter about 21 minutes after the trigger, and the optical source was clearly detected at the MASTER position \citep{2014GCN.15916....1X}. Nanshan’s observations were calibrated to the Mondy data by multiplying a normalisation factor, which is extracted by comparing the flux density at the same observation epoch.

\subsubsection{BOOTES-4 and all sky camera}
The 60\,cm Burst Observer and Optical Transient Exploring System (BOOTES)-4/MET robotic telescope in Lijiang (China) automatically responded to the trigger alert and started the first search in the clear filter at 13:23:08 UT, i.e., 36.8\,s after the \swift-BAT trigger. No optical transient was detected down to 18.3 mag (T0 +0.9 hr) in the first search. A series of automatic observations was conducted in \textit{g}, \textit{r}, \textit{i}, and \textit{z}-filters, providing upper limits on each search as summarised in Tab. \ref{tab:optical_obs}. The CASANDRA-4 \citep{2008SPIE.7019E..1VC} all-sky camera image obtained at the BOOTES-4 station in Lijiang (China) implies an upper limit for any optical flash simultaneous to the prompt $\gamma$-ray emission of R $>$ 7.5 mag.

\subsubsection{RATIR}
RATIR camera on the 1.5\,m Harold L. Johnson telescope at the Observatorio Astronómico Nacional on Sierra San Pedro Mártir in Baja California, Mexico began its observation of the field about 14.4 hours after the BAT trigger \citep{2014GCN.15937....1B, 2014GCN.15928....1B} and determined, the photometric redshift of $z_{phot}$ = 5.45$_{-0.2}^{+0.1}$ \citep{2014GCN.15928....1B}. The RATIR consists of four detectors, two optical and two infrared cameras, allowing four images of a source to be taken simultaneously, either in \textit{r,i,Z,J} or \textit{r,i,Y,H} \citep{2012SPIE.8453E..1OF, 2012SPIE.8453E..2SK, 2012SPIE.8444E..5LW, 2012SPIE.8446E..10B}. The images were reduced in near real-time using a Python-based automatic pipeline. Image alignment was conducted with astrometry.net \citep{2010AJ....139.1782L}, and image co-addition was achieved with Astromatic/SWarp \citep{2010ascl.soft10068B}. We use SExtractor \citep{1996A&AS..117..393B} to calculate photometry for individual science frames and mosaics with apertures ranging from 2 to 30 pixels in diameter, with an optical and NIR pixel scales of 0.32$''$ pixel$^{-1}$ and 0.3$''$ pixel$^{-1}$, respectively. Taking a weighted average of the flux in these apertures for all stars in a field, we construct an annular point-spread function. This point source photometry is then optimised by fitting the PSF to the annular flux values of each source. The \textit{r}, \textit{i}, and \textit{z} filters are calibrated to the SDSS DR9. We found that the RATIR and SDSS \textit{r}, \textit{i}, and \textit{z} filters agree to within 3\%, and the \textit{J} and \textit{H} filters are calibrated relative to the Two Micron All Sky Survey (2MASS; \citealt{2006AJ....131.1163S}). We use an empirical relation for \textit{Y} in terms of \textit{J} and \textit{H} magnitudes derived from the United Kingdom Infrared Telescope (UKIRT) Wide Field Camera observations (WFCAM; \citealt{2009MNRAS.394..675H}).

\section{Analysis and results} \label{sec:analysis}
\subsection{Prompt emission data reduction} \label{sec:pmt_lc}
The procedures for time-averaged and time-resolved spectral analysis during prompt emission are given in the Appendix, and results are summarised in Tab. \ref{tab:pmt_param}. In the following subsections, we provide the results obtained from the analysis process of the multi-band observations of \grb during the prompt and afterglow phase from \fermi, \swift missions, and the ground-based observatories listed above.

\begin{figure*}
\centering
\includegraphics[width=\linewidth]{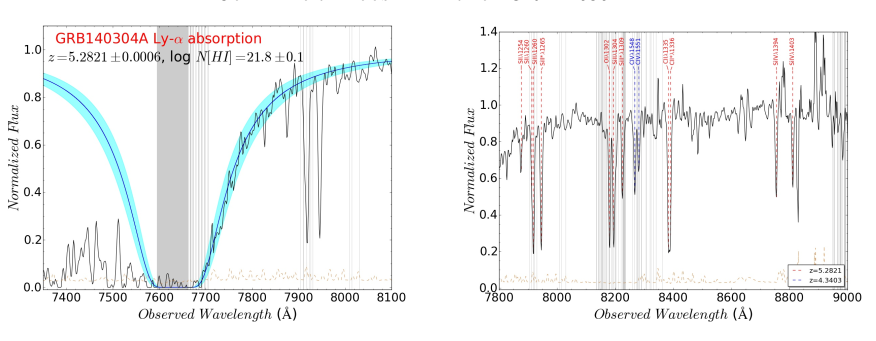}
\caption{In the left panel represents the absorbed L$_{y,\alpha}$ red damping wing is fitted with the Voigt profile. The solid cyan area represents the 68\% confidence interval. Similarly, the right panel represents the spectrum and absorption lines detected by the Gran Telescopio Canarias (GTC) on 04 March 2014, from the \grb afterglow.}
\label{fig:GTC_spectra}
\end{figure*}

\subsubsection{Multi-band prompt emission light curve} \label{sec:prompt_lc}
The multi-band light curve of \grb taken from \swift-BAT and \fermi-GBM is shown in Fig. \ref{fig:multiband_plc}. The upper two panels represent the light curve from \swift-BAT and the corresponding hardness ratio. The middle and lower panels represent the light curve from \fermi-GBM observations from NaI and BGO detectors and the corresponding hardness ratio. The \grb is a multi-pulse burst, where several small pulses overlap the two broad pulses. In the upper panel, the \swift-BAT light curve is plotted with magenta (15-25\,keV) and green (50-100\,keV). The hardness ratio is calculated by dividing the number of photons detected in the 50-100\,keV energy range by the number of photons in the 15-25\,keV energy range of \swift-BAT. It is clear from the upper two panels that during the \tninty duration, the counts in the 50-100\,keV energy band are higher than the counts in the 15-25\,keV energy band. The \swift-BAT observation hints that the \grb is a hard burst. Similarly, from \fermi-GBM observations, again, the number of photons in the energy range 50-300\,keV (red curve in the middle panel) is slightly higher than the number of photons in the energy range 8-50\,keV (green curve in the middle panel), which again confirms the hard nature of the \grb. The lower panel represents the light curve in the BGO energy range, where the yellow light curve is plotted in the energy range 1-40\,MeV and the blue light curve is in the energy range 0.2-1\,MeV. \grb hardly have detections in the 1-40\,MeV energy band, and most of the detections of this burst lie below the 1\,MeV range. When we compare the light curve from the upper to the lower panel of Fig. \ref{fig:multiband_plc}, we have noticed that the pulse width of \grb decreases with energy. The \swift-BAT observed broad pulses with a full width at half maximum (FWHM\footnote{calculated from Fig. \ref{fig:multiband_plc}, although not shown in the figure}) = 5.04\,s, which become narrower in the NaI energy range with an FWHM = 4.90\,s, and in the BGO observation, the pulse becomes further narrower with an FWHM = 2.42\,s. This property is commonly observed in multi-pulse GRBs, as mentioned in earlier studies \citep{2005ApJ...627..324N}.

\subsection{Afterglow data analysis}
The following sub-sections describe the process of data analysis utilised for $\gamma$-ray, X-ray and optical observations from various ground and space-based telescopes.

\subsubsection{BAT and XRT light curve} \label{sec:XRT_lc}
We extracted the mask-weighted unabsorbed \swift-BAT flux density curves at 10\,keV from the Burst Analyser page of UK Swift Science Data Centre \citep{eva07, eva09}. We have also included the bad BAT data points\footnote{Some BAT data points may be defined as bad, either because, for SNR-binned light curves, the bin never achieved a signal-to-noise ratio of 3 or because the count-to-ﬂux conversion factor had an uncertainty greater than two orders of magnitude}, in our light curve. The X-ray flux density light curve at 10\,keV taken from BAT and XRT observations is plotted in the top panel of Fig. \ref{fig:multiband_AGLC}. 

From Fig. \ref{fig:multiband_AGLC}, we see that both the BAT and XRT light curves consist of several peaks throughout the duration of the afterglow. The observed peak in the XRT light curve in the 100 - 1000 s range overlaps with the BAT light curve, suggesting their common origin. The XRT light curve also indicates that the central engine of the burst remains active at least up to this timescale ($\sim$10$^{5}$\,s), a detailed discussion about this is given in section \ref{sec:XRT_flares}.

We have calculated the {time of the peaks} in the observed \swift-XRT light curve using the \sw[find\_peaks] algorithm implemented in the \sw[SciPy] library. The \sw[find\_peaks] algorithm checks for each data point to determine whether it represents a peak by comparing it with neighbouring data points. With this method, we have obtained around 15 peaks in the \swift-XRT light curve, which further indicating that the erratic behaviour of the central engine strongly affects the prompt and afterglow emission phases. The inset of Fig. \ref{fig:multiband_AGLC} shows the time period over which the 4 brightest X-ray peaks occur. Further, the entire (including flare) X-ray light curve is fitted by a power-law function with two breaks. Based on the observed temporal indices, we assumed that the X-ray light curve consists of at least three components. The first one is a plateau with the decay index $\alpha_{X,1} = 0.24_{-0.11}^{+0.11}$, and the second component is a steep decay with the decay index $\alpha_{X,2} = 3.47_{+0.45}^{-0.36}$, and again followed by a plateau phase with $\alpha_{X,3} = 0.47_{-0.33}^{+0.42}$. The observed breaks are lies at $t_{b,X,1} = 1608.5_{-298.7}^{+246.7}$\,s and $t_{b,X,2} = 5924.1_{-1438.6}^{+6850.4}$\,s. \\

According to \cite{2015AdAst2015E..13G}, for flares caused by a reverse shock, the decay indices should be 3 or 2 in an ISM or \textit{Wind}-like medium, respectively, and these flares should dominate at lower energy bands such as NIR and optical. However, the observed flares are steep {(decay indices $<$ 3)} and simultaneous across all bands, which rules out the external reverse shock origin. Additionally, the steep decay following the plateau is consistent with an internal origin \citep{2007ApJ...665..599T}. 

All three phases are significantly affected by the overlapping nature of flares, possibly due to near-simultaneous emission from the engine active for a prolonged duration. The observed rise and decay indices of the flares, shown in the top panel of Fig. \ref{fig:multiband_AGLC}, are too steep to be explained by the external forward and reverse shock models. Therefore, our analysis suggests that for \grb, the emission observed up to 1600\,s (${1600}/{(1+z)} \sim 255$\,s, in rest frame) originates from an internal origin.

\subsubsection{Optical light curve} \label{sec:opt_lc}
To compare the features of the multi-wavelength observations of \grb, we have plotted optical observations together with the BAT and XRT observations in the top panel of Fig. \ref{fig:multiband_AGLC}. We have also plotted the multi-band optical observation from various ground-based optical telescopes listed in Tab. \ref{tab:optical_obs}. Similar to the X-ray light curve, the optical light curve of the \grb consists of several flares. The \textit{R}-band light curve consists of the highest number of observations of all the optical bands. Therefore, we utilised the \textit{R}-band light curve for the fitting process. The \textit{R}-band light curve exhibits a power-law decaying component that overlaps with several flares. The power-law index obtained from the \textit{R}-band light curve is $\alpha_{R} = 1.04 \pm 0.03$, though the fitting is performed without removing overlapping flares. The flares observed in the optical light curve are not exactly aligned but delayed with the positions of flares in the X-ray light curve, which is consistent with the positive lag observed in LGRBs. Further, the late optical light curve consists of 3 data points each in the \textit{r}, \textit{i}, and \textit{z} bands that are not consistent with the observed power-law decay in the optical light curve, though there are no earlier \textit{i} and \textit{z} points to confirm this claim. The deviation of the last three optical data points from the underlying power-law component is consistent with the late flare observed in the X-ray light curve. We have fitted a smoothly joined broken power-law function to \textit{r} band to determine the {peak time of optical flares}; for details, please refer to \citet{2007AA...469L..13M, 2025MNRAS.543.2404R}. Due to the scarcity of optical data during the late phase, exact calculation of the peak is not possible, but roughly lies at $t_{\rm p, opt} \sim 20000$\,s followed by a decay with index $\alpha_{d} \sim 1.7$ in the late-time \textit{r} band data.

\subsection{Spectral energy distributions (SEDs) of \grb} \label{sec:sed}
The optical-X-ray SEDs were created at four different epochs (SED1, SED2, SED3, and SED4) corresponding to the four bright peaks observed in the X-ray afterglow light curve. The SEDs are shown in the bottom panel of Fig. \ref{fig:multiband_AGLC} and corresponding epochs are mentioned in the first column of Tab. \ref{tab:X_idx}. For the given intervals, the X-ray spectra are retrieved from \swift's online repository\footnote{\url{https://www.swift.ac.uk/xrt_spectra/00590206/}} in 0.3-10 keV and combined with the corresponding optical observations as shown with the light shaded regions in green, red, blue, and yellow, respectively, for SED1 to SED4. For the first epoch or SED1, in addition to the optical and X-ray observations, we have also utilised the observations from the \swift-BAT and \fermi's Large Area Telescope (LAT; \citealt{2009ApJ...697.1071A}) as a limiting value.\\

The retrieved X-ray spectra in 0.3-10 keV, along with other broadband SEDs, are then fitted by a power-law function. The results of spectral/SED fittings are respectively listed in columns 2 and 3 of Tab. \ref{tab:X_idx}. The spectral slopes from X-ray in 0.3-10 keV are denoted by $\beta_{X}$, and those obtained from the joint fit of optical to X-rays are denoted by $\beta_{OX}$. For the SED1, the X-ray spectral slope is $\beta_{X} \sim 1.5$, which implies an electron spectral distribution index $p \sim 3$ ($2\beta_{X}$), considering X-ray emission is above the cooling frequency ($\nu_{c}$). At the same time, the $\beta_{X}-\beta_{OX}\sim 0.5$. This indicates that the $\nu_{c}$ lies between the optical and X-ray bands \citep{Sari_1998} around the epoch of SED1. For SED2, SED3 and SED4, derived values of both $\beta_{X}$ and $\beta_{OX}$ are consistent within the errors, and the resulting electron energy distribution index $p \sim 2\beta_{X}+1$ in the range 2.6-2.8, considering entire optical and X-ray emission lying below $\nu_{c}$. The obtained $p$-values are listed in column 4 of Tab. \ref{tab:X_idx}. This behaviour is consistent with a $\it Wind$-like circumburst medium, where the cooling break frequency increases with time and crosses the X-ray band after the epoch corresponding to the SED1 \citep{Sari_1998, 2013NewAR..57..141G}. Therefore, the observed afterglow evolution of \grb is broadly consistent with the \textit{Wind}-like medium, also found by \cite{2018ApJ...859..134L} using radio data.

\subsubsection{Steeper optical slope \(\beta_{o}\) during SED4 and estimated extinction} \label{sec:extinction}
The multi-band optical-NIR simultaneous observations from RATIR (\textit {r,i,z} bands) at 25,000\,s (SED4) exhibit a steep optical spectral slope \( \beta_{o} \approx 6 \). Using the relation between hydrogen column density and visual extinction from \citet{1995A&A...293..889P}, \(A_V = N_H / 1.8 \times 10^{21}\), we estimated the optical extinction value \( \rm A_V = 3.5~mag\) based on our measured \( N_H \) (see next section). Even after correcting for the above extinction value following the procedure from \cite{2024A&A...690A.373R}, the optical spectral slope remains significantly steeper than the corresponding X-ray spectral slope at a similar epoch. A higher extinction value of \(\rm A_V = 4 - 5~mag \) is required to reconcile the two indices falling in the same spectral regime as per applicable forward shock model predictions described above. The obtained A$_{V}$ values differ markedly from the predictions for high-$z$ GRBs \citep{2024A&A...690A.373R} and align with those observed in some of the dark GRBs at large distances \citep{2009AJ....138.1690P}. However, these $A_V$ corrections yield a consistent optical spectral slope \(\rm \beta_{o} \sim 0.9\) with optical frequencies lying below $\nu_{c}$ corresponding to an electron distribution index \( p \sim 2.6 \), as discussed above as a larger picture.

Also, the late-time decay index $\alpha_{d} \sim 1.7$ is consistent with the relation \(\alpha = \frac{3p - 1}{4}\) for a $\it Wind$-like medium \cite{Sari_1998}, implying an \(p \sim 2.6\), for the emission below $\nu_{c}$. These analysis results align well with those reported by \citet{2018ApJ...859..134L}. It is also noted that the SED4 created at $\sim$25,000\,s exhibited a significant deviation from those predicted for typical afterglows, though created during the flaring activity. Additionally, the \textit{i}-band observations used to construct the SED4 might also have been partly affected by flux losses due to the Lyman-$\alpha$ absorption effects as typically seen in Lyman-break galaxies \citep{2005MNRAS.359.1184S}. The non-detection of host to deeper limits \citep{2024ApJ...966..133S} and observed redder afterglow colours [(i-z)/(r-z) in the range 1.3 to 1.9 mag] demands a higher value of extinction in the burst direction (as discussed above) and suggests the source lies in a very faint galaxy where the drop in the \textit{i}-band flux due to the high redshift, characteristic of Lyman-break galaxies \citep{2024A&A...691A.240M, 2024ApJ...966..133S}.

\subsection{Redshift determination from GTC optical spectrum} \label{sec:GTC_redshift}
The increase in brightness level at optical-NIR frequencies during the late flaring activity (see Fig. \ref{fig:multiband_AGLC}) enabled us to find the spectroscopic redshift of \grb by triggering GTC. The combined spectrum exhibits the continuum only after 7750 \AA~with a clear discontinuity cut of L$_{y,\alpha}$, see left panel of Fig. \ref{fig:GTC_spectra}, which implies that \grb is a high redshift burst with \(z \sim 5\) \citep{2014GCN.15922....1J, 2014GCN.15936....1J}. We fitted the absorbed L$_{y,\alpha}$ red damping wing using the Voigt profile and found hydrogen column density as log(N$_{H,Opt}$/cm$^{2}$) = 21.8 $\pm$ 0.1. A series of absorption lines resulting from different species (S, Si, SiII$^{*}$, OI, C, and CII$^{*}$) are clearly detected at the same redshift of the L$_{y,\alpha}$ feature and confirmed that the afterglow is situated at \(z = 5.2821 \pm 0.0006\), see right panel of Fig. \ref{fig:GTC_spectra}, this redshift is utilised in the analysis throughout the paper. Also, at least one intervening CIV system is found at \(z = 4.3403 \pm 0.0001\). We excluded the SiII-$\lambda$1304 line in this computation since it looks blended with an unknown intervening feature, and no constraint on this component can be imposed in order to perform an accurate de-blending.

\begin{table}[]
\centering
\caption{Fit parameters parameters from optical to X-ray SED}
\begin{tabular}{|l|c|c|c|} \hline
Epoch (s) & $\beta_{X}$ & $\beta_{OX}$ & $p$ \\ \hline
SED1~(80-110) & 1.51$_{-0.23}^{+0.23}$ & 0.92$_{-0.03}^{+0.03}$ & 3.0 $\pm$ 0.4 \\
SED2~(275-600) & 0.69$_{-0.11}^{+0.11}$ & 0.80$_{-0.03}^{+0.03}$ & 2.6 $\pm$ 0.2 \\
SED3~(745-960) & 0.71$_{-0.15}^{+0.15}$ & 0.81$_{-0.03}^{+0.03}$ & 2.6 $\pm$ 0.3 \\
SED4~(18-26)$\times10^{3}$ & 0.77$_{-0.29}^{+0.29}$ & 0.91$_{-0.03}^{+0.02}$ & 2.8 $\pm$ 0.5 \\ \hline
\end{tabular}
\tablefoot{The optical to X-ray SED created at four epochs corresponding to peaks in the observed X-ray light curve. The $\beta_{X}$ spectral indices obtained from the fitting of X-ray spectra in 0.3 to 10\,keV fitted by an absorption Power-law. The $\beta_{OX}$ obtained from fitting the optical to X-ray SED using a simple power-law. The $p$ values are obtained using the closure relation in the \textit{Wind}-like medium.}
\label{tab:X_idx}
\end{table}

\begin{table}[]
\centering
\caption{EW measurements for the systems detected on the \grb afterglow spectrum.}
\begin{tabular}{|c|c|c|c|c|} \hline
Feature & Wavelength & $z$ & EW & eEW \\ \hline
SII$\lambda$1254 & 7875.5 & 5.2813 & 0.23 & 0.04 \\
SII$\lambda$1260 & 7913.4 & 5.2829 & 0.42 & 0.01 \\
SiII$\lambda$1260 & 7919.3 & 5.2831 & 0.85 & 0.03 \\
SiII$^{*}$$\lambda$1265 & 7945.8 & 5.2825 & 1.03 & 0.03 \\
SiII$\lambda$1304 &-&-&-&-\\
SiII$^{*}$$\lambda$1309 & 8224.6 & 5.2818 & 0.49 & 0.01 \\
CII$\lambda$1334 & 8383.5 & 5.282 & 0.86 & 0.02 \\
CII$^{*}$$\lambda$1336 & 8390.1 & 5.2814 & 0.85 & 0.02 \\
SiIV$\lambda$1394 & 8755 & 5.2816 & 0.65 & 0.04 \\
SiIV$\lambda$1403 & 8812.1 & 5.2819 & 0.5 & 0.05 \\
CIV$\lambda$1548 & 8267.9 & 4.3403 & 0.64 & 0.02 \\
CIV$\lambda$1551 & 8281.6 & 4.3403 & 0.5 & 0.02 \\ \hline
\end{tabular}
\label{tab:EW}
\end{table}

\begin{figure}
\centering
\includegraphics[width=\columnwidth]{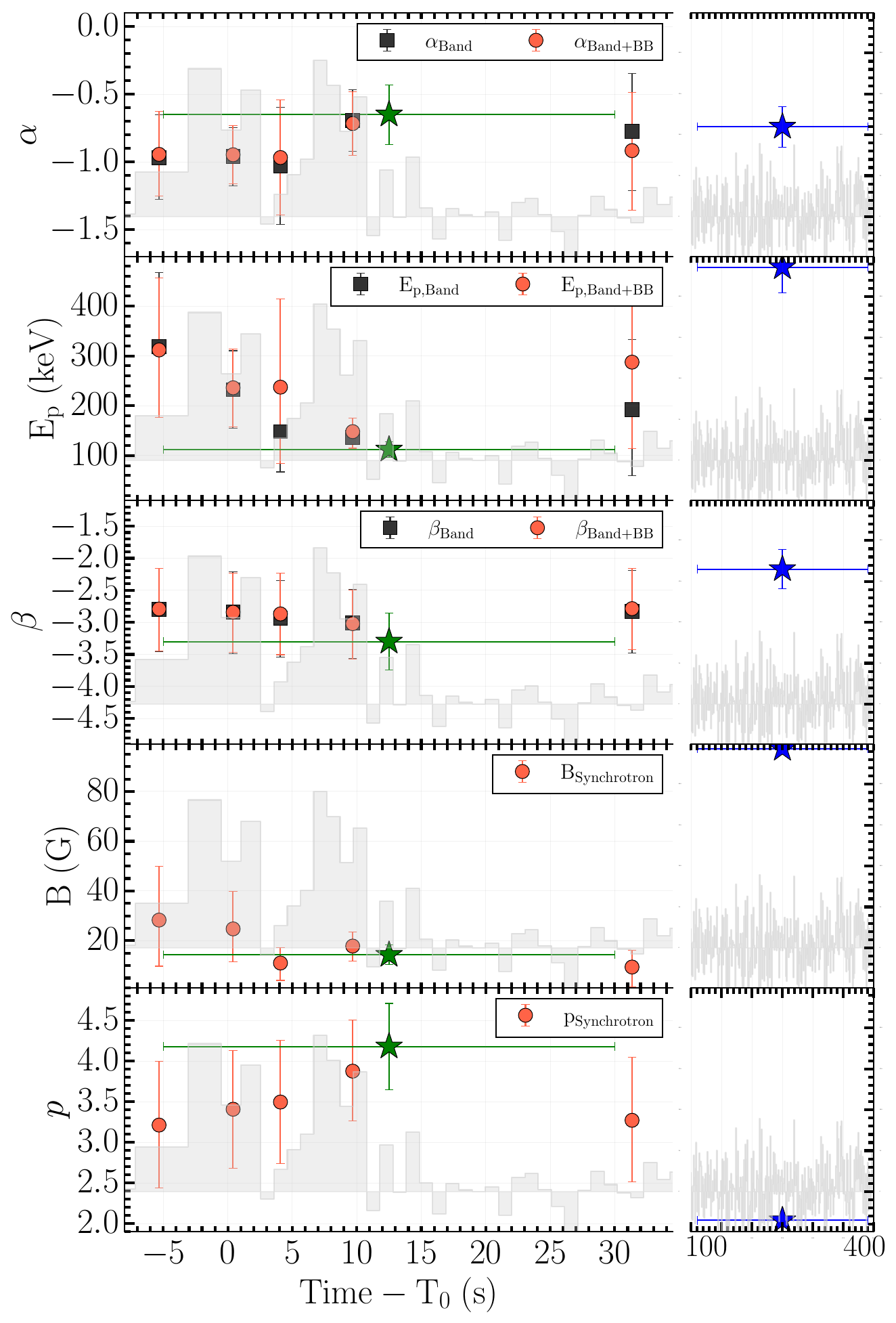}
\caption{The evolution of prompt emission spectral parameters obtained from the fitting of \fermi-GBM observations of \grb. From top to bottom, the panel 1, the evolution of the low energy spectral index ($\alpha$) is shown, obtained from the fitting of the \sw[Band] and \sw[Band+Blackbody] functions. Similarly, panels 2 and 3, respectively, show the evolution of \Ep and $\beta$. Panels 4 and 5 represent the evolution of magnetic field strength (B) and electron energy distribution index ($p$) obtained from the fitting of the physical synchrotron model \citep{2020NatAs...4..174B}. The green and blue stars represent the parameter obtained from time-integrated spectral fitting, respectively, between [-5 - 30]\,s and [100 - 400]\,s.}
\label{fig:prompt_params}
\end{figure}

\section{Discussion} \label{sec:discussion}
In this section, we elaborate major results found from the present analysis of the prompt emission and afterglow observations of \grb.

\subsection{Evolution of prompt emission parameters} \label{sec:Pspec_evo}
The temporal evolution of spectral parameters during the prompt emission phase serves as a key diagnostic for understanding the underlying physical emission mechanisms. In the following subsection, we examine the evolution of these spectral parameters derived from the prompt emission phase.

\subsubsection{Time-integrated analysis}
The rest frame peak energy \(E_{p, rest} = 116 \times (1+z) = 728\)\,keV obtained from the fitting of time integrated prompt emission spectra with band function, see Fig. \ref{fig:prompt_params} and Tab. \ref{tab:pmt_param}, the isotropic energy \(E_{\gamma, iso} = 5.7 \times 10^{52}\) erg of \grb is consistent with the Amati and Yonetoku correlation within the LGRB population \citep{Amati, Yonetoku_2004}. Additionally, considering its continuous emission from the prompt and afterglow phases, the burst’s overall duration and the spectral behaviour (see Fig. \ref{fig:Ep_t90}) align more closely with those of LGRBs. However, as shown in Fig. \ref{fig:bat_xrt_lcs}, having late flares in the afterglow light curves seems common to the majority of high-$z$ GRBs. Furthermore, as discussed in the next section, an observed positive lag in the prompt emission light curve of the burst confirms its origin as LGRB. \\

\subsubsection{Time-resolved analysis}
The prompt emission spectral parameters obtained from the fitting of several empirical and physical models are shown in Fig. \ref{fig:prompt_params} and Tab. \ref{tab:pmt_param}. These parameters do not remain constant and often evolve with time, providing constraints on the underlying emission mechanisms \citep{1986ApJ...301..213N, 2019ApJ...884..109L, 2021MNRAS.505.4086G}. The widely accepted mechanism for the prompt emission is the synchrotron emission from the cooling population of accelerated electrons from internal shock or magnetic reconnection, respectively, in the baryon-dominated or Poynting-flux-dominated outflows \citep{magnetic_reconnection, 2018NatAs...2...69Z}. The peak energy \Ep of the \(\nu F_{\nu}\) spectrum has broadly four types of evolution during the prompt emission phase: i.e. (i) Hard-to-soft evolution in which \Ep decreases steadily with time; (ii) Intensity-tracking evolution, in which \Ep follows the light curve of the burst; (iii) Soft-to-hard evolution in which \Ep increases steadily with time and (iv) the chaotic or random evolution \citep{1986ApJ...301..213N, 1994ApJ...422..260K, 1983Natur.306..451G}. 

In our analysis of the prompt emission spectra of \grb (see section Appendix\ref{sec:pmt_analysis}), we have found that the \Ep has a hard-to-soft evolution, and at the same time \(\alpha\) does not evolve much during the prompt emission and remains within the prediction of synchrotron emission \citep{2002ApJ...581.1248P}. The synchrotron model is also favoured, as some of the time-resolved bins are best fit by the physical synchrotron model \citep{2020NatAs...4..174B}. The evolution of the magnetic field strength (\(B\)) and the electron energy distribution index (\(p\)) reveals that \(B\) decreases over time while \(p\) seems to increase, but evolution is not clear due to large error bars. This further supports a progressive softening or hard-to-soft evolution of the prompt emission spectra, which is commonly interpreted as a consequence of synchrotron cooling or adiabatic expansion of the emitting plasma \citep{2021A&A...656A.134G}. Therefore, for \grb, the hard-to-soft evolution is possibly due to the synchrotron emission from the cooling population of accelerated electrons in the external magnetic field. As the fireball expanded, the strength of internal shock and the magnetic field reduced over time, which reflected as a hard-to-soft evolution. \\

Further, the time-resolved spectral analysis of the other two high redshift GRBs (GRB 220521A \& GRB 240218A having good dataset from \fermi-GBM) has shown diverse results (see Fig. \ref{fig:GRB22_24}), i.e., hard to soft evolution for GRB 220521A and GRB 240218A exhibited intensity tracking evolution. Therefore, our analysis suggests that hard-to-soft evolution is not typical for high-$z$ GRBs, indicating that the jet dissipation process and the origin of the prompt emission are independent of the burst redshift; however, more such observations are needed for a robust conclusion about possible progenitors/underlying physical mechanisms of high-$z$ GRBs.

\subsection{Spectral lags in prompt emission light curve of high-$z$ GRBs} \label{sec:HZ_lag}

In case of GRBs, spectral lag refers to the time delay observed between the arrival of photons in different energy bands for current detectors/instruments \citep{1986ApJ...301..213N, 2005ApJ...627..324N}. Earlier results have shown that the SGRBs usually have negligible or small negative lags, i.e, low-energy photons arrive earlier than high-energy photons \citep{2006MNRAS.367.1751Y, 2014AstL...40..235M}. A negative delay can also be the result of the superposition of several pulses that make up the time profile \citep{2014AstL...40..235M}. The majority of LGRBs show positive lag (high-energy photons arrive earlier than low-energy photons) due to the dominant softening trend in their spectra \citep{2000ApJ...534..248N, 2007ApJS..169...62H, 2021ApJ...922...34C}. However, in some cases, the transition from positive to negative lag is observed \citep{2017ApJ...842..115W, 2025A&A...697A.161M}. The observed median lag in LGRBs can range from $\sim$50 ms to a few hundred ms, depending on the selected energy bands and burst properties \citep{2006MNRAS.367.1751Y}. A negative lag could be due to complex jet structure, such as the variation of the Lorentz factor with the jet, magnetic reconnection in the outflow, Inverse Compton scattering, or Lorentz invariance due to which high-energy photons arrive later \cite{2016ApJ...825...97U, 2021ApJ...922...34C}. The positive spectral lag is assumed to be related to the curvature effect of the jet, i.e., photons from the outer edges of the jet arrive later and remain softer due to lower Doppler boosting \citep{2002ApJ...578..290R}. Another reason for the positive spectral lag is the hard-to-soft parameter evolution during the prompt emission phase \citep{2003ApJ...594..385K, 2011AN....332...92P, 2012MNRAS.419..614U, 2021ApJ...922...34C, 2023ApJ...942...67L}.\\

\begin{figure}[t]
\centering
\includegraphics[width=\columnwidth]{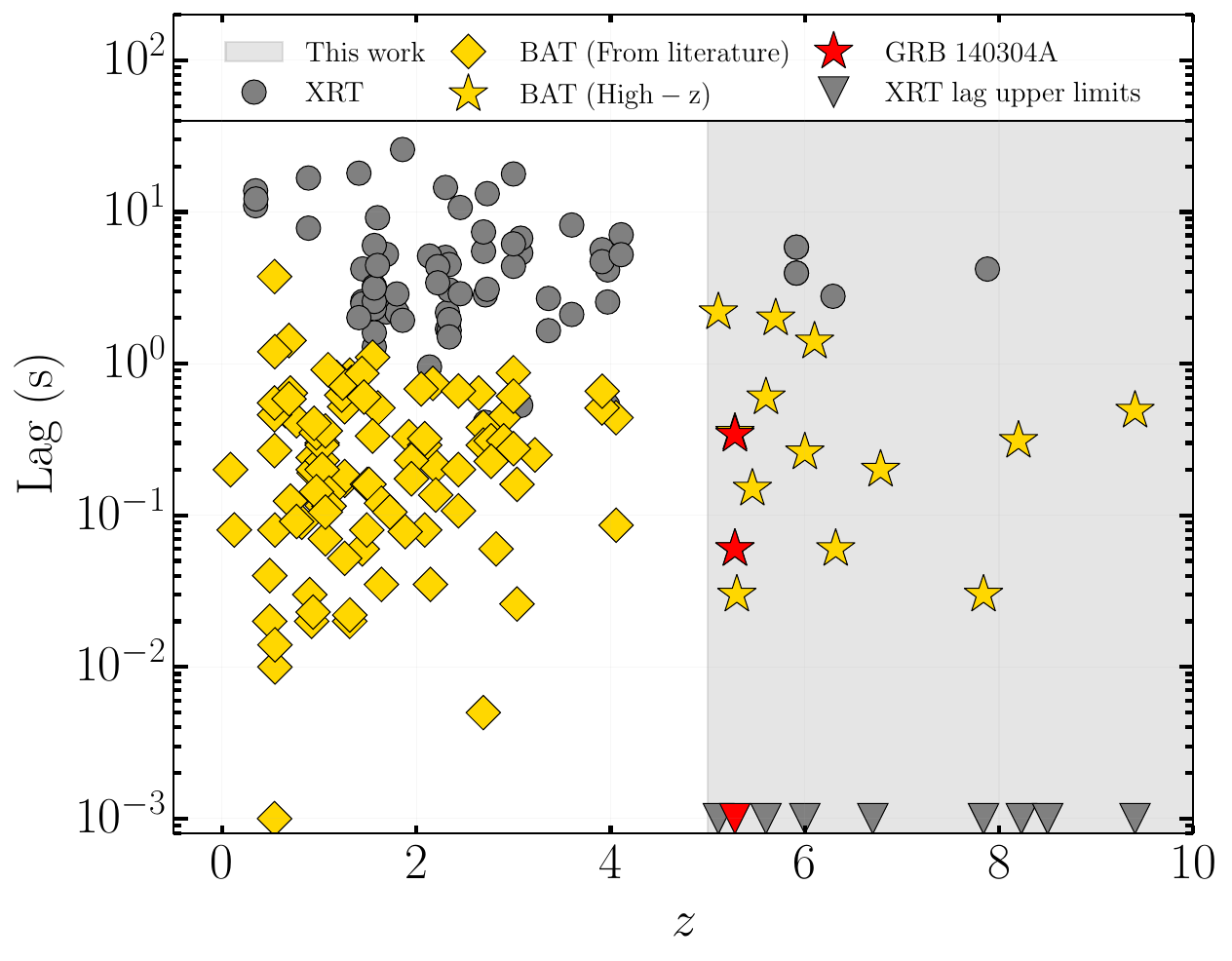}
\caption{Spectral lags obtained from the cross-correlation function using the prompt emission light curve in two energy bands \(50-100\)\,keV and \(15-25\)\,keV of \swift-BAT (yellow legends) and \(0.3-1.5\)\,keV and \(1.5-10\)\,keV of \swift-XRT from the early afterglow phase (grey legends). The yellow diamonds for \(z<5\) represent the spectral lag taken from the literature \cite{2012PASP..124..297L, 2012MNRAS.419..614U}. The grey circles for \(z<5\) represent the lag in \swift-XRT taken from \cite{2021ApJ...922...34C}. In this figure, the spectral lag values for high-$z$ GRBs with \(z>5\) calculated as part of the present analysis are shown with yellow stars and lie within the shaded region. The downward triangles are plotted to show that no spectral lag is constrained in the XRT light curves of some high-$z$ GRBs.}
\label{fig:lag_high_z}
\end{figure}

\begin{figure}[t]
\centering
\includegraphics[width=\linewidth]{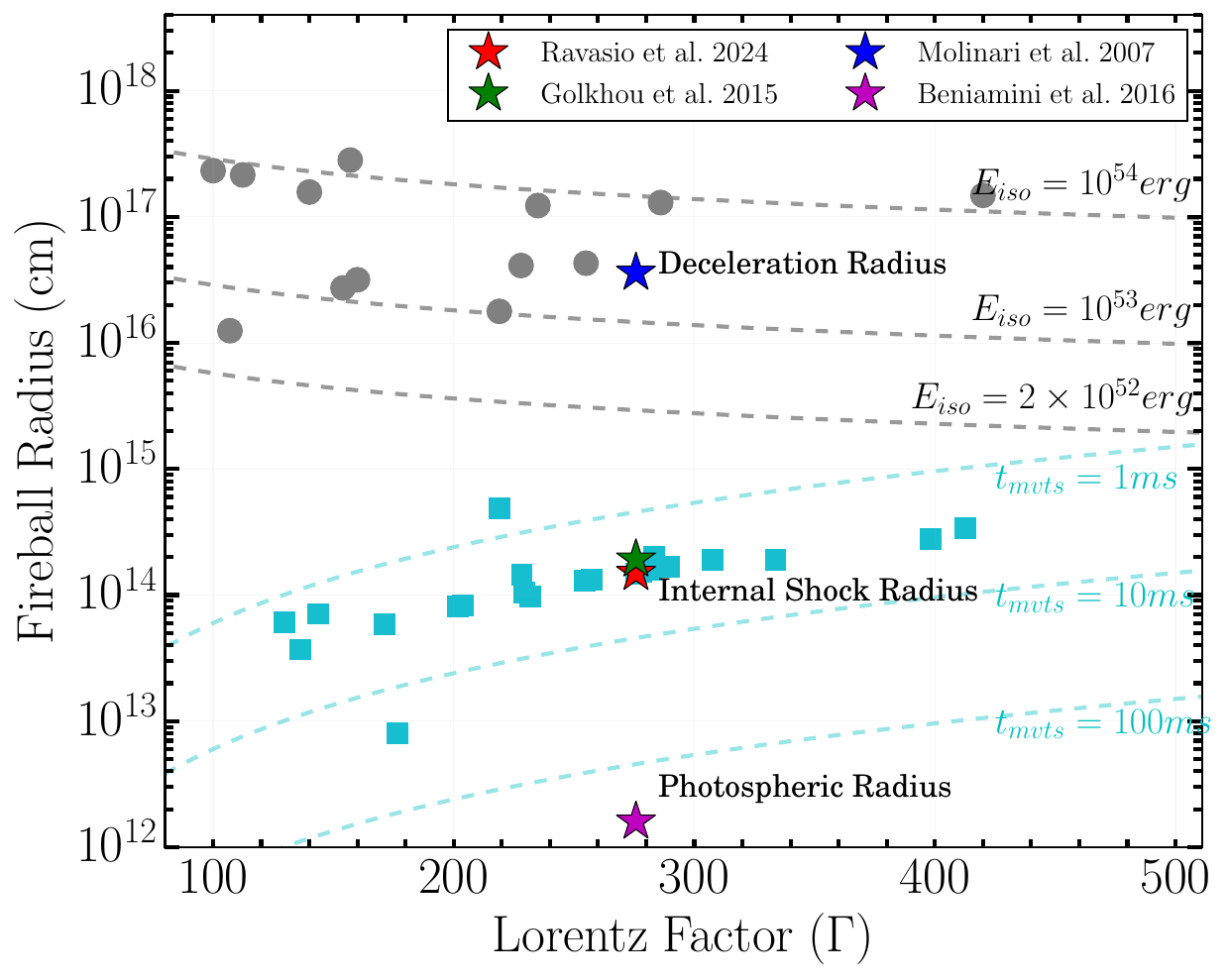}
\caption{Represents the fireball radius calculated for \grb at different epochs from early prompt emission to late afterglow phases. The different methods used to estimate the fireball radius are mentioned in the legend. The cyan data points (filled squares) represent the fireball radii estimated during the prompt emission phase \citep{2024A&A...685A.166R} and the grey data points (filled circles) represent the fireball/deceleration radius from \cite{2021MNRAS.505.4086G} for comparison. The cyan and grey dashed lines are plotted using equations 3 and 4 from \cite{2025A&A...702A..95M}.}
\label{fig:IS_radii}
\end{figure}

Following the procedure of \cite{2012MNRAS.419..614U}, we have calculated the spectral lag between [50-100]\,keV - [15-25]\,keV of \swift-BAT and in [1.5-10]\,keV - [0.3-1.5]\,keV of \swift-XRT, in the observed frame. To calculate the spectral lag, we have used a cross-correlation function (CCF, \citealt{1997ApJ...486..928B, 2023ApJ...942...67L, 20130348M}) following the methodology of \cite{2012PASP..124..297L, 2022MNRAS.511.1694G}. For \grb, positive lags \(0.34\pm0.10\)\,s and \(0.06\pm0.06\)\,s are constrained for the first and second peak of the \swift-BAT light curve, and no lag is found in the \swift-XRT light curve. The spectral lag calculated for \grb and the corresponding cross-correlation function is shown in Fig. \ref{fig:CC_lag}. Similarly, we have also calculated the spectral lag in the \swift-BAT light curve high redshift GRBs with \(z > 5\), as shown in Fig. \ref{fig:lag_high_z} and Tab. \ref{tab:master_table}. In Fig. \ref{fig:lag_high_z}, to compare the lag in high-$z$ GRBs with other well-studied LGRBs, we also utilise the data from \cite{2012PASP..124..297L, 2012MNRAS.419..614U}. We have also included the data points from the \cite{2021ApJ...922...34C}, which represent the spectral lag in 48 GRBs with observed multiple flares in the \swift-XRT band.

As we have discussed in section \ref{sec:Pspec_evo}, \Ep has four types of evolution, which have different relations with the observed spectral lag in the prompt emission phase \citep{2003ApJ...594..385K, 2011AN....332...92P, 2023ApJ...942...67L}. In the case of hard-to-soft evolution, the spectral peak energy \Ep shifts from higher to lower energies over time. As a result, high-energy photons are emitted earlier and low-energy photons arrive later, leading to a positive spectral lag. In section \ref{sec:Pspec_evo}, our time-resolved spectral analysis revealed a hard-to-soft evolution of \Ep, which might account for the observed positive lag for \grb, supporting the findings of \cite{2012MNRAS.419..614U}. 

Further, our analysis found no evidence to support a redshift dependence/Lorentz invariance of the observed spectral lag in GRBs as discussed in the literature \citep{2016_Mochkovitch, 2016ApJ...825...97U, 2023ApJ...942...67L}.

\subsection{Evolution of GRB fireball} \label{sec:FB_evo}
Both in collapsar and binary merger scenarios of GRBs, stellar remnants generate a highly relativistic outflow, generally termed a fireball. This fireball is extremely dense and opaque initially, and undergoes rapid acceleration during its evolution \citep{1990ApJ...365L..55S, 1993MNRAS.263..861P, 1999PhR...314..575P, 2004RvMP...76.1143P}. At this stage, thermal emission is also expected from the photosphere (the outer layer where the fireball becomes transparent to radiation), having radius typically given as \(R_{\text{ph}} \sim \frac{L \sigma_T}{8 \pi m_p c^3 \Gamma^3}\) cm \citep{1986ApJ...308L..47G, 2000ApJ...530..292M, 2016MNRAS.457L.108B}. So, for \grb, (E$_{\gamma, iso} = 5.7 \times 10^{52}$\,erg and Lorentz factor \(\Gamma \sim 276 = 182 \times E_{\gamma,iso}^{-0.25}\) calculated using the relation from \cite{2010ApJ...725.2209L}), the photospheric radius $R_{ph}$ is $\sim$ 6 \(\times 10^{12}\)\,cm.\\

As the fireball continues to expand over time, radiation begins to escape from it. During this stage, internal shocks likely occur within the fireball, producing non-thermal emission. We have used two methods to estimate the fireball radii where these internal shocks take place: Firstly, \cite{2015ApJ...811...93G, Veres_2023, 2025A&A...702A..95M} proposed that the internal radius can be constrained using the relation \( R \sim 6 \times 10^{14} [\Gamma/100]^{2} [{t_{\rm mvts}/1\,s}]\) cm, where $t_{\rm mvts}$ is the minimum variability timescale. We have constrained the minimum variability timescale ($t_{mvts} \sim 1.1$\,s) by binning the GBM light using the Bayesian block binning technique in \sw[3ML]. The one-half of the width of the shortest bin is used as the minimum variability \citep{2013ApJ...764..167S, 2018ApJ...864..163V}. Using the relation and $t_{\rm mvts}$, we have determined that for \grb the radius of the internal shock emission is \( \sim 1.9 \times 10^{14}\)\,cm. Whereas \cite{2024A&A...685A.166R} constrains the fireball radius by considering the cutoff energy of high-energy photons due to the pair production process. In the case of \grb, the \fermi-LAT did not detect any high-energy $\gamma$-ray emission. The prompt emission light curve in the bottom panel of Fig. \ref{fig:multiband_plc}, plotted from the BGO detector data, reveals that the counts in 0.2 - 1 MeV are higher and only a little emission beyond 1 MeV. In addition to this, the SED is plotted in the Fig. \ref{fig:PSED}, revealing almost no photons above 10 MeV. Using this energy as a cutoff, we constrain the fireball radius at \( \sim 1.5 \times 10^{14}\)\,cm. Both the minimum variability and the cutoff-energy approach yield similar estimates for the radius at which internal shocks occur within the fireball. The values obtained for \grb are consistent with theoretical models that suggest gamma-ray emission dominates either from internal shocks or from magnetic dissipation processes occurring at large distances from the photosphere \citep{lyutikov2001}.

After the prompt emission phase, the fireball begins to interact with the surrounding medium. When the fireball collects enough mass, it loses a considerable part of its initial kinetic energy, i.e., also known as the deceleration radius (\(R_{dec}\)). Beyond this, the Lorentz factor of the fireball decreases with radius \(R \propto \Gamma^{-2-2\beta}\) following the self-similar solution \citep{1976PhFl...19.1130B}. This phenomenon, also known as the "onset of the afterglow", is usually identified as a smooth bump in the early optical light curves, where the peak time \(t_{p}\) of the afterglow light curves corresponds to the \(R_{dec}\). \cite{2007AA...469L..13M} provided the relation to calculate the \(R_{dec}\) using the early bump due to "onset of the afterglow", given as \( R_{dec} = {2ct_{p}\Gamma_{0}^{2}}/{(1+z)}\) cm. For \grb, the optical light curve does not show a smooth bump, which suggests that the onset of the afterglow is not well covered. Therefore, we have used the first data point (i.e., $t_p$ $<$ 83\,s) as a lower limit for the peak time constraining $\Gamma_{0}>675$ and \(R_{dec}\) $>$ \( 3.6 \times 10^{17}\) cm.

The calculated values of fireball radii at different epochs are presented in Fig. \ref{fig:IS_radii}, along with a sample of constraints for internal and external shock radii for a subset of well-studied GRBs \citep{2021MNRAS.505.4086G, 2024A&A...685A.166R}. In Fig. \ref{fig:IS_radii}, for fixed values of isotropic gamma-ray energy, the fireball deceleration radius shows a decreasing trend (Grey dashed lines in Fig. \ref{fig:IS_radii}) with the Lorentz factor. On the other hand, the internal shock radius shows an increasing trend (blue dashed lines in Fig. \ref{fig:IS_radii}) with the Lorentz factor, depending on the observed minimum variability timescales within the prompt emission light curve \citep{2025A&A...702A..95M}. The fireball radii obtained for \grb using various methods are consistent with these model predictions. 

It is also relevant to mention that recently published work by \cite{2026MNRAS.545f1985M}, {which discusses the radii of pop-III stars that may be the progenitors of high-$z$ GRBs, predict photospheric radius, internal shock and deceleration radius, consistent with the radii we find above (Fig. \ref{fig:IS_radii}), considering timescales such as engine time/bore time and duration} of bursts in case of LGRBs \citep{2011ApJ...739L..55B, Mizuta_2013}. However, more high-$z$ GRBs are required to be studied to systematically understand whether high-$z$ GRBs have Pop-III progenitors.

Our early optical observations up to the given limiting brightness and the non-detection of the onset feature (along with typical $E_{iso,\gamma}$ and higher $\Gamma > 675$) suggest that fireball deceleration may not have occurred within the observed time scale {(at least up to 10$^5$\,s)}, providing internal shocks as a possible explanation for the observed multi-band flares in \grb, though detailed calculation of such a possibility is beyond the scope of this paper.

\begin{figure}[h]
\centering
\includegraphics[width=0.95\columnwidth]{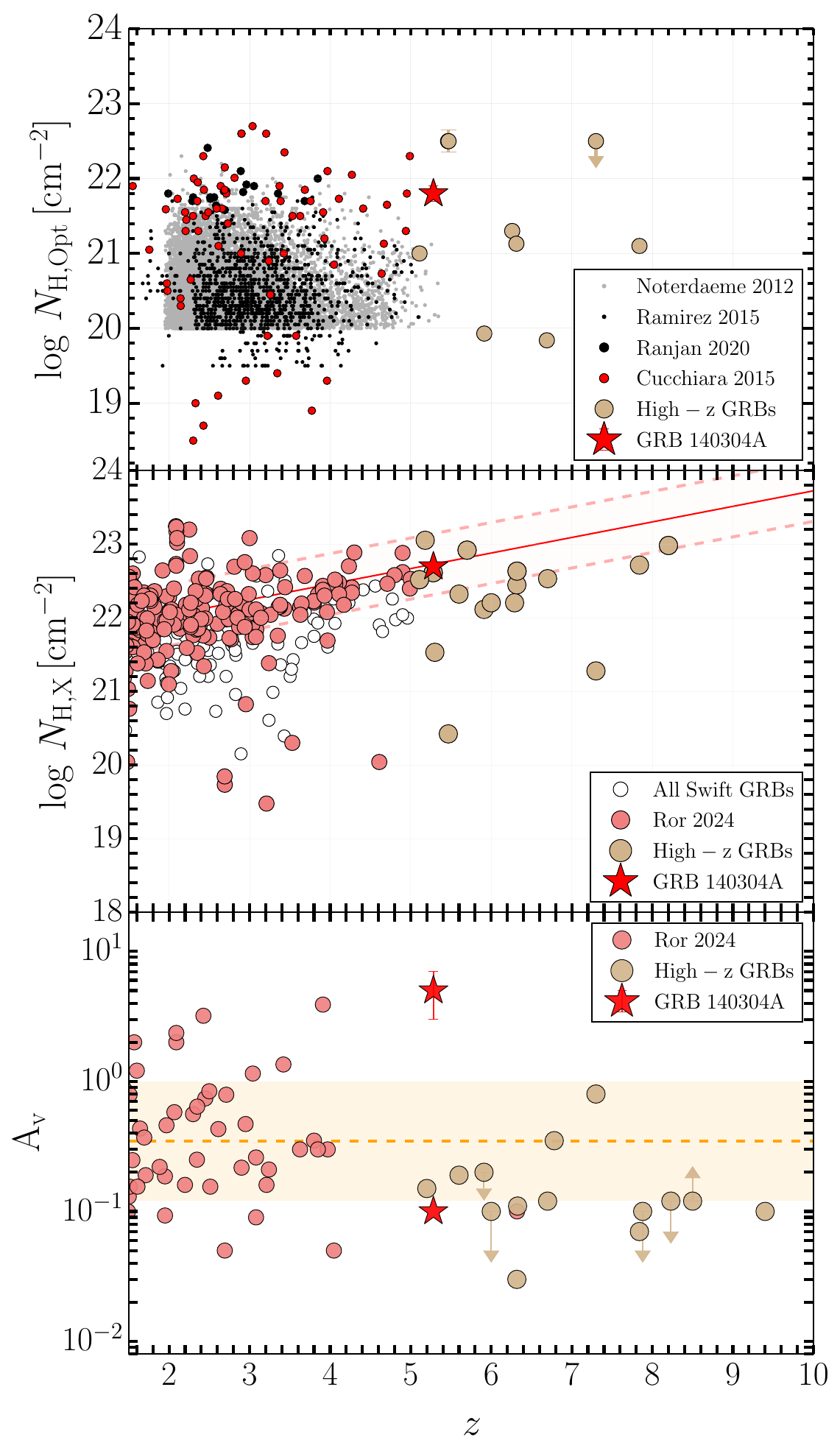}
\caption{Upper panel: Comparison of column density of \grb with the GRB-DLA compilation by \cite{2015ApJ...804...51C}, and the QSO-DLA by \cite{2016MNRAS.456.4488S}. QSO-DLA information is complemented with the log(N$_{H,Opt}$) $\geq$ 20 DLAs from the SDSS sample \citep{2012A&A...547L...1N, 2020A&A...633A.125R}. Middle panel: The hydrogen column density log(N$_{H,X}$) obtained from fitting of \swift-XRT spectra is plotted as a function of redshift, including data from redshift known GRBs (red circles) compiled by \cite{2024ApJ...971..163R, 2024Galax..12...51A}. High-$z$ GRBs are shown (golden circles, from Tab. \ref{tab:master_table}), with \grb highlighted by a red star. The red solid line represents a straight-line fit with a slight positive slope of 0.21$\pm$0.02, and the shaded region indicates the 1$\sigma$ spread of the fit. Lower panel: the host extinction in V band compiled for LGRBs and the high-$z$ GRBs (see Tab. \ref{tab:master_table}) from the published paper \citep{2010ApJ...720.1513K, 2010MNRAS.401.2773S, 2011A&A...525A.109D, 2014ApJS..213...15W, 2017MNRAS.467.1795L, 2018MNRAS.479.1542Z, 2022ApJ...940...57N, 2022ApJ...940...53S}. The two stars represent extinction for \grb, we have calculated the extinction $\rm A_V$ = 4-6 mag by fitting SED, and a lower value of extinction $\rm A_V$ = 0.1 mag is given in \cite{2018ApJ...859..134L}. The dashed line gives the mean value of $\rm A_{V}$ and the shaded region represents the $1 \sigma$ deviation.}
\label{fig:Column_Density}
\end{figure}

\subsection{Near-simultaneous, multi-wavelength late-time flares of \grb} \label{sec:XRT_flares}

As shown in Fig. \ref{fig:bat_xrt_lcs}, \grb is among the longest high-$z$ bursts detected by \swift-BAT in softer energy channels, with emission extending up to $\sim$700 s in the observed frame (corresponding to $\sim$110 s in the rest frame). Although its duration does not exceed the conventional threshold of $\sim 1000$\,s for ULGRBs, \grb stands out having most prolonged episodes of superimposed flaring activity observed by \swift-BAT. This, along with the accompanying \swift-XRT superimposed flaring and subsequent decay behaviour, distinguishes \grb as a unique high-$z$ event exhibiting one of the longest central engine activity at a range of frequencies.\\

Further, it is clear from Fig. \ref{fig:multiband_AGLC}, the observed X-ray emission from \grb does not show typical features of a canonical X-ray light curve \citep{2006ApJ...642..389N, 2006ApJ...647.1213O, 2006ApJ...642..354Z} and has an extended overlapped signature of irregular central engine activity witnessed by both XRT and BAT instruments. Our analysis in section \ref{sec:analysis} revealed that the X-ray light curve has a rapid decay after prompt emission, a plateau, and more than three subsequent X-ray flares (occurring at about 161\,s, 374\,s, and 824\,s at early epochs and a late-time flare around 25000\,s) hence decay indices different than those exhibited typically in case of other well-studied GRBs \citep{2009MNRAS.395..490O, 2025MNRAS.543.2404R}. We also studied the evolution pattern of spectral photon indices during the flares in the \swift-XRT band\footnote{The evolution of $\Gamma_{XRT}$ given on the burst analyser page of \grb \url{https://www.swift.ac.uk/burst_analyser/00590206/}}. The observed X-ray flares exhibit a distinctive pattern of spectral evolution. A hardening of the XRT-band photon index $\Gamma_{XRT}$ during the rising phase, followed by a softening of the index $\Gamma_{XRT}$ during the falling phase, though possibilities discussed for the origin of such early flares in GRBs
can not be ruled out \citep{2025arXiv251207731D}. This characteristic behaviour is typically observed in X-ray flares of other flaring sources \citep{2025ApJ...981..173A}.

Similar to X-ray emission, optical flares followed by a normal decay of the early light curve are clearly shown at 500\,s, 1100\,s, and, most probably, at 1600\,s and 20000\,s for \grb. Lastly, the $\gamma$-ray light curve from \swift-BAT consists of flares corresponding to at least two pairs of flares in the X-ray light curve, though the second flare in the BAT data is not statistically significant ($\sim$2 $\sigma$). Nevertheless, by assuming a Gaussian profile, we can estimate the true position of the first peak. The results of the analysis are presented in Tab. \ref{tab:G_fit} and shown in the upper panel of Fig. \ref{fig:multiband_AGLC}, where the positions of Gaussian peaks representing each flare in different energy bands are tabulated, showing epoch of flares being delayed at lower energies. Also, it is found that superimposed features/flares observed in canonical XRT light curves of some of the well-studied bursts \citep{2006ApJ...642..389N, 2006ApJ...647.1213O, 2006ApJ...642..354Z, 2008Natur.455..183R} during plateau to late-time decay phases can not necessarily be interpreted as prompt emission origin. However, currently popular models are unable to interpret these observed features in a unified manner and require more detailed theoretical investigation.

By and large, the observed afterglow light curve of \grb along with the present sample of other high-$z$ bursts having superimposed multiple flares/variabilities \citep{2006ApJ...642..389N, 2012ApJ...758...27L} are not significantly different from those afterglow patterns/features seen in many of the well-studied GRBs \citep{2025arXiv251207731D, 2025MNRAS.543.2404R}. The expected number of more such high-$z$ observations in the near future by \EP/\svom/others will be able to decipher about the possible progenitors in a better manner.

\begin{table*}[]
\renewcommand{\arraystretch}{1.5}
\centering
\caption{The prompt emission spectral and temporal properties of high-$z$ GRBs are compared with \grb}
\scriptsize
\begin{tabular}{|p{1.5cm}|r|c|c|c|c|c|c|c|l|c|} \hline
 GRB & BAT/GBM/ & \tninty &  $z$ & Time interval &  Lag &  $\Gamma_{BAT}$ & \Ep & E$_{\rm \gamma,iso}$ & Opt-NIR &  A$_{V}$\\
&  Konus &  (s)& &  (s) &  (s) & & (keV) &  (erg) &  Detection & (mag)\\ \hline
050502B\textsuperscript{1} & Yes/No/No & 17.5 $\pm$ 0.2 & 5.2 & & & -1.6 $\pm$ 0.1 & 89$\pm$30 & 3.8$\pm$0.7 & Yes & 0.15$\pm 0.07$\\
050904A\textsuperscript{18,19,23,26} & Yes/No/No & 182$\pm$16.3 &6.29(S) & 20-40 & 1.4 & & $>$150 & 31.74$\pm$1.22 & Yes & nan\\ 
060522A\textsuperscript{2,19,24} & Yes/No/No & 69.1$\pm$5.9 & 5.11(S) & -5-25 & 2.18 & -1.54$\pm$0.15 & 73.3$\pm$16.8 & 5.28$\pm$0.51 & Yes & 0\\ 
060927A\textsuperscript{3,19} & Yes/No/No & 22.4$\pm$1.19 &5.6(S)& -3-3/3-7/10-30 & 0.15/0.14/-0.88& -0.9$\pm$0.4 & 72$\pm$39 & 7.7$\pm$2.8 & Yes& $\sim$0.19 \\ 
080913A\textsuperscript{4,19, 24} & Yes/No/Yes & 8$\pm$1 &6.7 (S)& -10-10 & -0.06& -0.82$\pm$0.53 & 121$\pm$39 & 7.0$\pm$0.4& Yes & 0.12 $\pm$ 0.03 \\ 
090423A\textsuperscript{5,19} & Yes/Yes/No & 10.3$\pm$1.06 &8.23(S)& -10-15 & 0.31& -1.81$\pm$0.09 & 53.2$\pm$5.8 & 6.39$\pm$0.37& Yes & $<$0.46\\ 
090429B\textsuperscript{6,19} & Yes/No/No & 6$\pm$1 &9.4(P)& -10-10 & 0.49 & -1.89$\pm$0.12 & 45.3$\pm$5.5 & 4.31$\pm$0.36& Yes & 0.1 $\pm$ 0.02\\ 
090709A\textsuperscript{21} & Yes/No/No & 89$\pm$3 &8.5 (P)& 10-20/20-50 &0.6/0.1 & -1.06$\pm$0.12 & 439$\pm$58 & 31$\pm$0.36& Yes & $>$2 \\ 
100905A\textsuperscript{7} & Yes/No/No & 3.4$\pm$0.504 &7.88(P)& -10-10 & -1.02 & -1.09$\pm$0.19 & 127.0$\pm$45.1 & 1.53$\pm$0.19& Yes & < 0.5\\ 
120521C\textsuperscript{8} & Yes/No/No & 27.1$\pm$4.34 &6.0(P)& -10-20 & 0.26 & -1.66$\pm$0.11 & 103.0$\pm$37.6 & 19 $\pm$ 8& Yes & $<$0.05 \\ 
120923A\textsuperscript{9} & Yes/No/No & 26.1$\pm$6.82 &7.84 (P)& -10-40 & 0.03 & -0.29$\pm$0.24 & 44.5$\pm$7.0 & 4.8$\pm$1.6& Yes & $\sim$0.07 \\ 
130606A\textsuperscript{10} & Yes/No/Yes & 277$\pm$19.6 &5.91 (S)& -10-10/150-158 & -0.12/0.13 & -1.52$\pm$0.12 & 150.0$\pm$72.2 & 16.81$\pm$1.28& Yes & $<$0.2 \\
 &  &  & & /158-168 & /0 &  &  & & & \\
131227A\textsuperscript{11} & Yes/No/No & 18$\pm$1.55 &5.3 (S)& -10-20 & 0.03 & & & 4.15$\pm$0.31& Yes & nan \\ 
140304A\textsuperscript{12} & Yes/Yes/No & 14.8$\pm$1.4 & 5.282(S)& -5-5/5-15& 0.34/0.06& -1.44$\pm$0.05 & 116$\pm$18 & 5.72$\pm$0.28& Yes & $\sim$5 \\ 
140515A\textsuperscript{13, 25} & Yes/No/No & 23.4$\pm$2.04 &6.32(S)& & & -1.78$\pm$0.12 & 56.4$\pm$9.9 & 5.8$\pm$0.6& Yes & 0.11 $\pm$ 0.02\\ 
201221A\textsuperscript{14} & Yes/{TS}/Yes & 44.3$\pm$5.95 &5.7(S)& 10-20&1.99& -1.4$\pm$0.15 & 113.0$\pm$36.4 & 10.58$\pm$0.96& No & nan \\ 
210905A\textsuperscript{15} & Yes/{TS}/Yes & 778$\pm$388 & 6.318(S)& -30-20&0.06 & -0.66$\pm$0.19 & 144$\pm$56 & 127$\pm$20 & Yes & $\sim$0.03 \\ 
220521A\textsuperscript{16} & Yes/Yes/No & 13.5$\pm$2.69 & 5.6(S)& -5-3/3-16 & 0.61/0.08& -1.97$\pm$0.18 & 48.8$\pm$40.6 & 4.32$\pm$0.44 & Yes & nan \\
240218A\textsuperscript{17} & Yes/Yes/Yes & 66.9$\pm$11.3 & 6.78 (S) & -2 - 20 & 0.2& -1.00$\pm$0.32 & 293$\pm$150 & 54$\pm$14& Yes & $\sim$0.35 \\ \hline
250314A\textsuperscript{22, 27} & SVOM & 11$\pm$3 & 7.3 (S)& & & -1.05$\pm$0.22 & 77$\pm$25 & 4.65$\pm$1.13& Yes & $\sim$0.8 \\ \hline
\end{tabular}
\tablefoot{Column 1 lists the high-$z$ GRBs, 2 lists detecting instruments, 3 lists \tninty duration, 4 lists redshifts, where (S) and (P) stand for spectroscopic and photometric redshifts, 5 lists the time interval over which spectral lag is calculated, 6 lists the observed value of Lag, 7 lists the BAT photon index, 8 lists the \Ep in the BAT energy range, 9 lists the isotropic energy release in multiple of $10^{52}$ erg in 15-150\,keV energy band. For GRB 250314A, all the values are derived from the SVOM mission. Column 10 lists whether there is a detection of the optical afterglow and the host. All quantities are in observed frames. A few GRBs are not detected by \fermi but are found in target search (TS). Values listed are taken from the references given below, otherwise driven from the BAT catalogue.}
\tablebib{[1] \citealt{2011A&A...526A.154A}, [2] \citealt{2006GCN..5153....1K} \label{060522A}, [3] \citealt{2007ApJ...669....1R} \label{060927A}, [4] \citealt{2009ApJ...693.1610G} \label{080913A}, [5] \citealt{2009ApJ...703.1696Z, 2009Natur.461.1258S} \label{090423A}, [6] \citealt{2011ApJ...736....7C} \label{090429B}, [7] \citealt{2010GCN.11218....1B, 2018_Bolmer} \label{100905A}, [8] \citealt{2012GCN.13333....1M, 2014ApJ...781....1L, 2017MNRAS.466.4558Y} \label{120521C}, [9] \citealt{2018ApJ...865..107T}, \label{120923A} [10] \citealt{2015_Hartoog, 2017MNRAS.466.4558Y} \label{130606A}, [11] \citealt{2013GCN.15623....1D} \label{131227A}, [12] Present work \label{140304A}, [13] \citealt{Chornock2014GRB1A, 2014GCN.16284....1S} \label{140515A}, [14] \citealt{2020GCN.29100....1M} \label{201221A}, [15] \citealt{Rossi_2022} \label{210905A}, [16] \citealt{2022GCN.32090....1L} \label{220521A}, [17] \citealt{2024GCN.35761....1B, 2025A&A...695A.239B} \label{240218A}, [18] \citealt{2016ApJ...825..135M}, [19] \citealt{2012ApJ...754...46T}, [20] \citealt{2008GCN..8256....1P}, [21] \citealt{2010AJ....140..224C}, [22] \citealt{2025A&A...704L...7C}, [23] \citealt{2007ApJ...665..102B}, [24] \citealt{2012A&A...542A.103B}, [25] \citealt{2015A&A...581A..86M}, [26] \citealt{2005_Tagliaferri, 2006Natur.440..184K}, [27] \citealt{2025A&A...704L...8L}.}
\label{tab:master_table}
\end{table*}

\subsection{Hydrogen column density measurements and physical interpretation}
In Section \ref{sec:GTC_redshift}, by fitting the absorbed L$_{y,\alpha}$ red damping wing with a Voigt profile, we obtain hydrogen column density log(N$_{H,Opt}$/cm$^{2}$) = 21.8 $\pm$ 0.1 for \grb, towards the higher end in the sample of high-$z$ bursts. For instance, our earlier study of high redshift burst GRB 130606A at $z = 5.91$ yielded a rather low log(N$_{H,Opt}$/cm$^{2}$) = $19.85 \pm 0.15\,\mathrm{cm}^{-2}$ \citep{20135631C}. Comparison of the hydrogen column density of \grb is made with those determined for other high-$z$ bursts/GRB damped L$_{y,\alpha}$ systems (GRB-DLA), and quasar DLA sample (QSO-DLA) as presented in the upper panel of Fig. \ref{fig:Column_Density}. We chose the QSO-DLA compilation to avoid any false samples in the SDSS DLA catalogue due to the statistics \citep{2012A&A...547L...1N, 2020A&A...633A.125R, 2016MNRAS.456.4488S}. We include in the upper panel of Fig. \ref{fig:Column_Density} the SDSS sample having log(N$_{H,Opt}$/cm$^{2}$) $\geq$ 21.7 DLAs in order to remark the detection of the other high column density systems, observed towards QSO lines of sight. The finding of a range of hydrogen column densities for the \grb along with other bursts is evidence that the GRB and QSO-DLA samples are drawn from different populations. We also measured the equivalent widths (EW) of the lines by fitting them with a Gaussian profile and computing the sum over the line model. This approximation is valid in the low resolution of the GTC/OSIRIS spectrum, as the convolution of the instrumental profile with the actual Voigt profile makes the absorption remain approximately Gaussian. Results are presented in Tab. \ref{tab:EW}. 

Further, we have also compared the hydrogen column density ($\log N_{{H,X}}$) obtained from fitting the \swift-XRT spectra of \grb to the large sample of GRBs from \cite{2024ApJ...971..163R} and high-$z$ GRBs, as shown in the middle panel of Fig. \ref{fig:Column_Density}. This plot reveals a clear trend of increasing $\log N_{{H,X}}$ with redshift \citep{2010MNRAS.402.2429C} having positive slope values similar to those published in literature \cite{2024ApJ...971..163R} and reference therein. High-$z$ GRBs (depicted in gold colour) generally exhibit higher $N_{\rm {H,X}}$ values compared to lower-redshift events, indicating denser surrounding environments in the early Universe. Notably, \grb (highlighted by the red star) stands out with one of the highest $N_{\mathrm{H,X}}$ measurements at $z \sim 5.282$, supporting the link between extreme column densities and high-$z$ GRBs. This pattern suggests that high-$z$ GRBs may preferentially occur in regions of substantial gas content, consistent with the predictions for environments hosting early massive Pop-III stars \citep{2015JHEAp...7...35S, 2024Galax..12...51A}. 

However, the host $A_{V}$ values compiled for high-$z$ GRBs, as shown in the lower panel of Fig.~\ref{fig:Column_Density} (\citealt{2022ApJ...940...57N, 2022ApJ...940...53S, 2024ApJ...971..163R} and reference therein), typically exhibit lower extinction values compared to other LGRBs. This contrasts with the typical correlation between $\log N_{{H,X}}$ and $A_{V}$ for the Galaxy, QSOs and other galaxies \citep{1995A&A...293..889P, 2024Galax..12...51A}. The high $N_{\mathrm{H,X}}$ (pointing to denser environments) and low $A_{V}$ (dust and metal-poor environments) values at high-$z$ challenge dust production processes or suggest different dust characteristics compared to the nearby universe, indicating distinct chemical evolution and physical conditions at play in the early universe. This atypical behaviour clearly indicates the special character of host galaxies at high-$z$ and is also indicative of special properties, such as the local metallicity and dust within the ambient media near the explosion site \citep{2018_Bolmer, 2024Galax..12...51A}. Given the limited sample of such high-$z$ GRBs, other important factors, such as selection biases (e.g., high X-ray column densities and low dust and metal content at high-$z$), cannot be ruled out. More such examples in the near future from SVOM/EP (in X-ray and $\gamma$-ray) and \textit{JWST} (NIR/Optical) will help to understand this least understood aspect of star-formation/evolution processes at high-$z$.

\subsection{Comparison of \grb with other GRBs at high redshift ($z$ $>$ 5)} \label{sec:HZ_comp}
\begin{figure*}
\centering
\includegraphics[width=0.9\columnwidth]{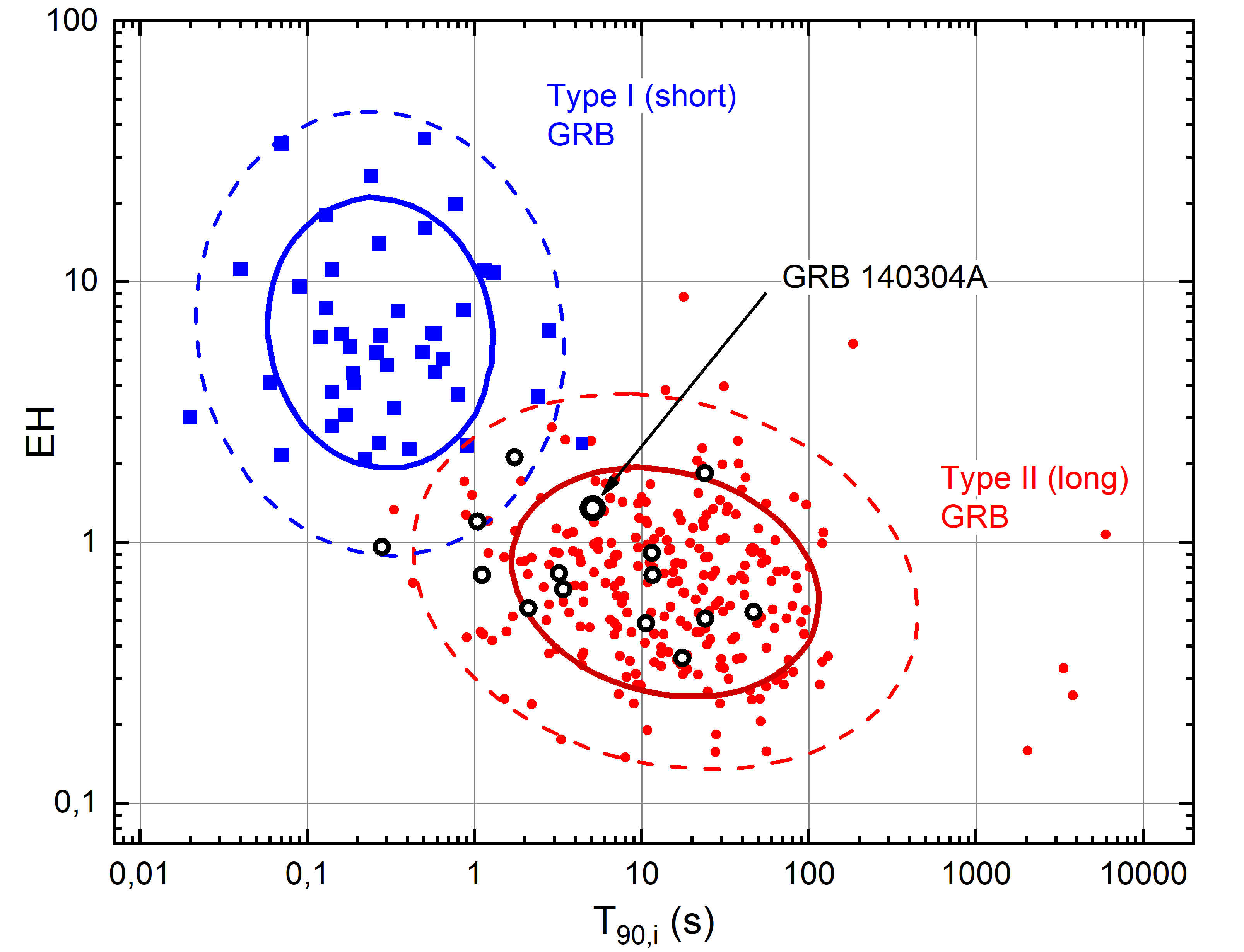}
\includegraphics[width=\columnwidth]{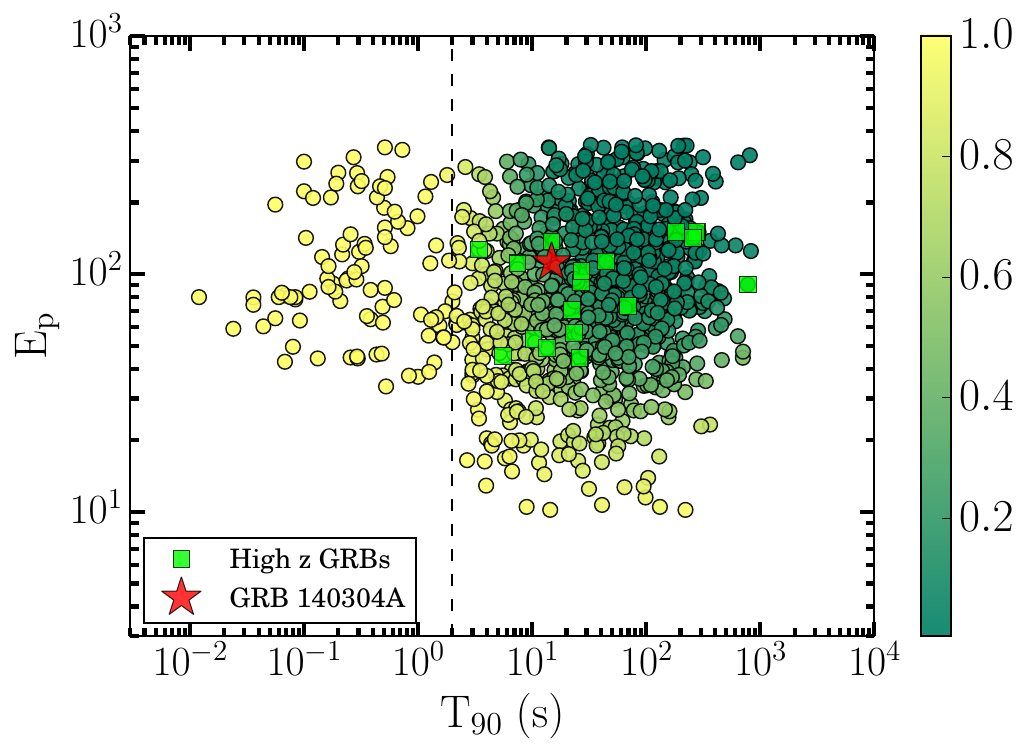}
\includegraphics[width=\columnwidth]{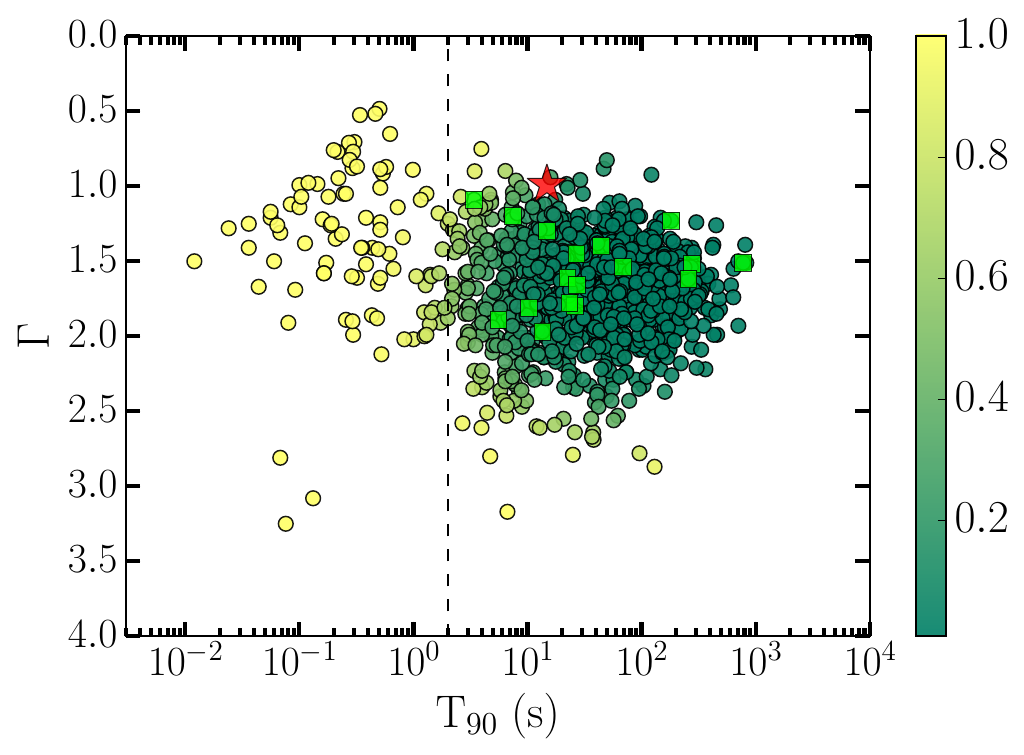}
\includegraphics[width=\columnwidth]{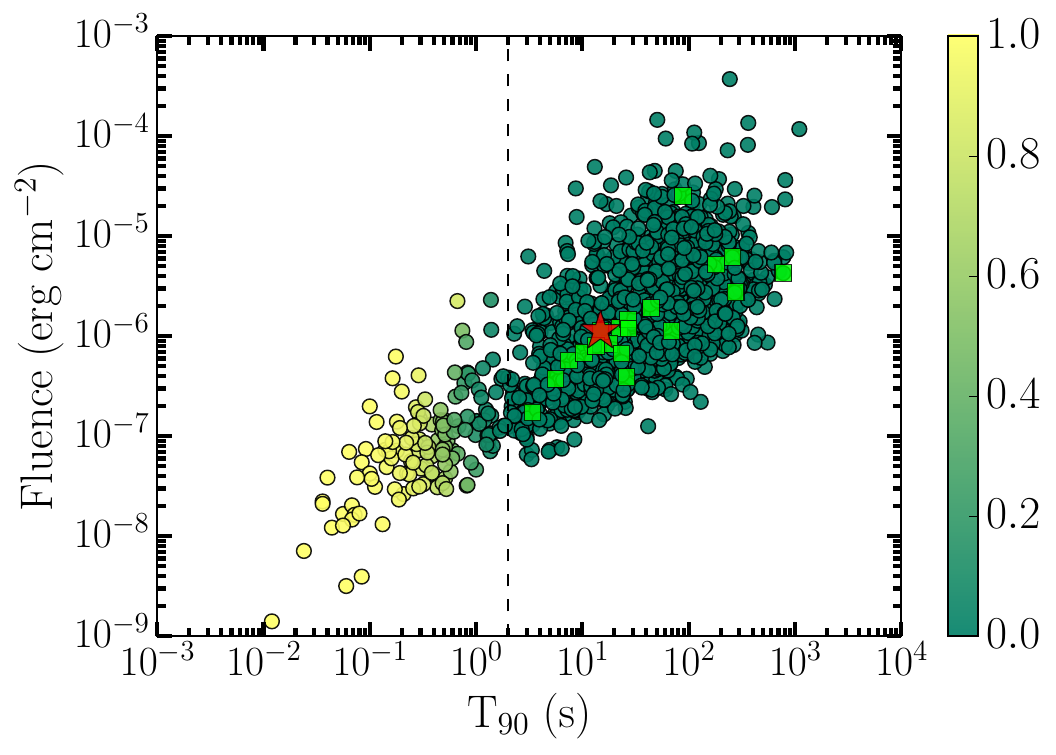}
\includegraphics[width=\columnwidth]{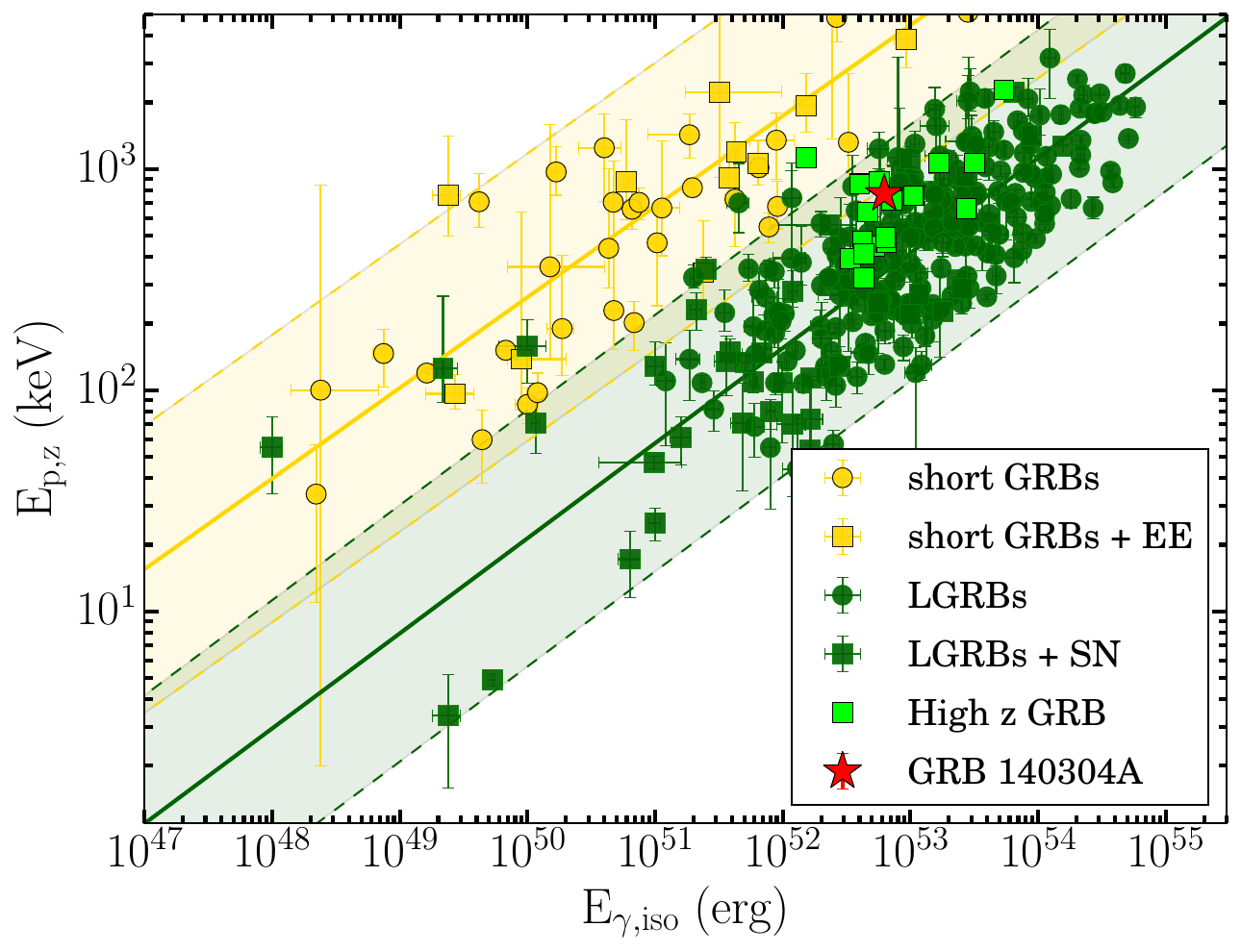}
\includegraphics[width=\columnwidth]{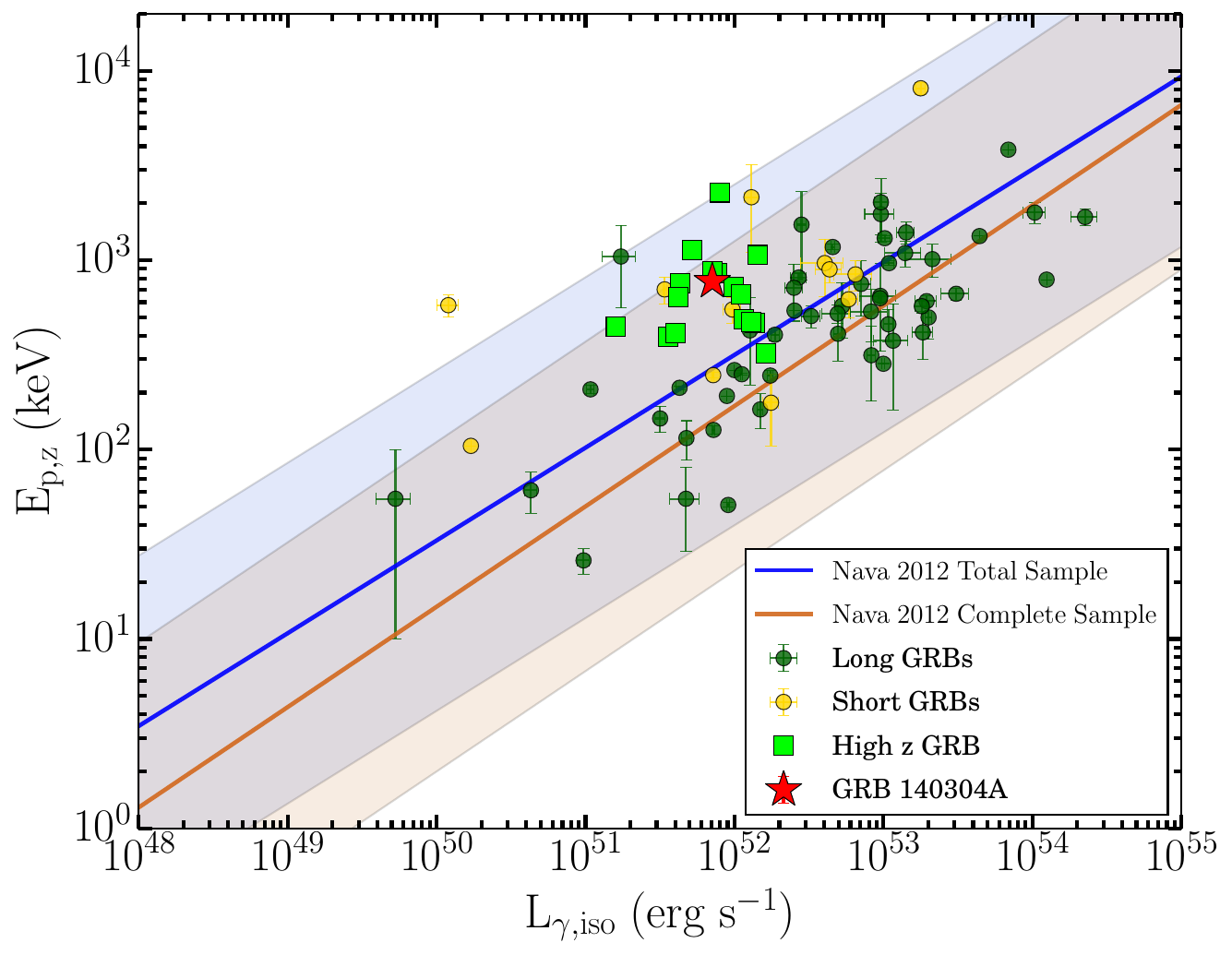}
\caption{Upper left: \grb in the combination of $E_p$ and $E_{iso}$ vs. \tninty \citep{minaev2020}. Upper right: \grb is shown in the \Ep-\tninty plane, along with other GRBs at high redshifts ($z>5$). \Ep $\sim 88$ of the burst is calculated from fitting a Band function to the Joint XRT-BAT spectra. Middle left: \grb in the $\Gamma$-\tninty plane. $\Gamma$ of the burst is calculated from the Joint spectral fitting of XRT and BAT observations. Middle right: \grb is shown in the Fluence-\tninty plane. The observed fluence of the burst is taken from the \swift-BAT webpage \citep{eva09}. In the lower-left panel, the high redshift GRBs are shown in the Amati plot, and the lower right similarly represents the high-$z$ GRBs in the Yonetoku plane.}
\label{fig:Ep_t90}
\end{figure*}

We have compiled 20 GRBs detected by \swift-BAT (including one from SVOM recently) observed at redshift $>5$ \citep{2006ApJ...638L..71B, 2009ApJ...693.1610G, 2009Natur.461.1258S, 2011ApJ...736....7C, 2018ApJ...865..107T, 2025A&A...704L...7C}, listed in Tab. \ref{tab:master_table}. Additionally, GRB 060116 was suggested to lie at \(z\sim6.60\), but only poorly constrained using photometric observations \citep{2024A&A...686A..56K}, we exclude this from our sample. The observed X-ray and {$\gamma$-ray} light curves for a nearly complete sample of high redshift GRBs are shown in Fig. \ref{fig:bat_xrt_lcs}. Comparison of prompt emission properties, such as Hardness Ratio (HR), spectral peak energy (\Ep), spectral photon index ($\Gamma_{BAT}$), fluence, \tninty of \grb and other high-$z$ GRBs along with the complete sample of GRBs detected with \swift is shown in Fig. \ref{fig:Ep_t90}. In the upper left panel of Fig. \ref{fig:Ep_t90}, we have calculated the HR using the \swift-BAT fluence in the energy range 50-100\,keV divided by fluence in 15-25\,keV using the values given in the \swift-BAT catalogue \citep{eva07, eva09}. In the HR-\tninty plot, the two distinct groups corresponding to the populations of LGRBs and SGRBs are shown separated by the vertical dotted line at 2\,s. It is clear from Fig. \ref{fig:Ep_t90} that the \grb has one of the highest HR values in the observed frames among the GRBs with \(z > 5\). This indicates that the prompt emission of \grb has relatively harder spectra compared to other high-$z$ GRBs. The hard nature of other two high-$z$ GRBs (GRB 080913 and GRB 090423A) has also been discussed by \cite{2009ApJ...703.1696Z}. Similarly, in the spectral peak energy-duration plane (\Ep-\tninty) and the photon index-duration plots ($\Gamma_{BAT}$-\tninty), \grb has one of the highest values of \Ep and also has the hardest photon index relative to other GRBs with \(z > 5\). We fitted a Gaussian mixture model to the given sample and calculated the probability of each GRB belonging to one of the groups (SGRBs or LGRBs). The probabilities of any GRBs belonging to the SGRBs class are shown in the colour bar in Fig. \ref{fig:Ep_t90}. Although \grb has hard prompt emission spectra, this GRB is only moderately bright, as shown in the fluence-\tninty plot. When we transformed the observed peak energy into a rest frame, we found that the rest-frame \Ep, isotropic energy (E$_{\gamma, iso}$), and isotropic luminosity (L$_{\gamma, iso}$) released during the prompt emission are consistent with the Amati \citep{Amati, 2012MNRAS.421.1256N} and Yonetoku \citep{Yonetoku_2004} planes (see lower most panels Fig. \ref{fig:Ep_t90}). In the Amati and Yonetoku planes, the observed $E_{p,z}$, $E_{\gamma, \, iso}$, and L$_{\gamma, iso}$ of the \grb are consistent with the general population of LGRBs. Our analysis suggests that even at high redshift, the \grb possesses the properties of typical LGRBs but has relatively harder prompt emission spectra than other GRBs with \(z>5\). These findings align with previous studies, suggesting that the prompt emission properties of high-$z$ GRBs are similar to those of typical LGRBs \citep{2009Natur.461.1258S, 2015JHEAp...7...35S, 2016ApJ...825..135M, 2017MNRAS.467.2476M, 2025A&A...695A.239B}.

\section{Conclusions} \label{sec:conclusion}
We present the multi-band photometric and spectroscopic observations of \grb (at \(z = 5.282\), exploded when the Universe was only 8\% of its current age), spanning optical to $\gamma$-rays. Our analysis reveals that towards the lower BAT energies, the observed frame duration of \grb is longest among the high-$z$ sample, however, the reported T$_{90}$ duration in the 15-350\,keV channel is $\sim15$\,s \citep{2014GCN.15927....1B}. The rest frame duration (\(T_{90, rest} \sim 2.3\,s\)) of \grb lies on the boundary of the SGRBs and LGRB, and the temporal and spectral properties of the burst during both prompt and afterglow phases are consistent with the LGRB population. Our analysis revealed that the prompt emission spectral peak energy \Ep and magnetic field strength B have a hard-to-soft evolution. The rest-frame \swift-BAT light curve displays a positive spectral lag, indicating energy-dependent photon arrival times during the prompt emission phase. The hard-to-soft evolution of \Ep also accounts for the positive lag observed in \swift-BAT light curves. The low energy spectral index $\alpha$ obtained from the \sw[Band] function did not show any significant evolution and remains consistent within the predictions of synchrotron emission from the slow or fast cooling population of accelerating electrons.\\

Throughout the afterglow phase, multiple flares are observed. {Flares or bumps in the light curve are not uncommon, see e.g. \cite{2018IJMPD..2744012M}}. Our analysis reveals a rarely observed morphological correspondence between the observed optical, X-ray, and probably $\gamma$-ray. There are systematic time delays among peak times of the flares in the three different energies, however, the physical origin of spectral lags over the wide frequency range needs to be understood further in terms of unified theoretical models \citep{2012MNRAS.419..614U, 2016_Mochkovitch, 2016ApJ...825...97U}.

The X-ray afterglow light curve of \grb consists of a plateau followed by a steep decay, consistent with the internal origin of the observed X-ray emission (including three bright flares) at least up to break time \citep{2007ApJ...665..599T}. Additionally, during the afterglow phase, the X-ray photon index ($\Gamma_{XRT}$) shows an anti-tracking behaviour with the X-ray light curve. This could be one of the reasons for the non-detection of a spectral lag within the \swift-XRT band, especially if lag is related to the spectral evolution. \\

The GTC spectral analysis and multi-wavelength SED fitting reveal an excess of neutral hydrogen column density and dust extinction along the line of sight, suggesting that \grb occurs in a dusty, \textit{Wind}-like medium. Alternatively, \grb could be hosted by a very faint i-drop galaxy characterised by properties similar to those of star-forming Lyman-break galaxies at high redshift \citep{2005MNRAS.359.1184S, 2024A&A...691A.240M, 2024ApJ...966..133S}.\\ 

The observed properties of \grb helped to study the comparison of observed properties collectively for a nearly complete sample of bursts observed at $z>5$ till date, including the one observed by \svom \citep{proceedingswei, 2025A&A...704L...8L} recently, extending insights into their physical origin, emission mechanisms of this rare subset of bursts.\\

Finally, our analysis indicates that flares/features observed in the limited sample of high-$z$ bursts, including \grb, are also common in other well-studied GRBs. More such observations would be required in the \EP, \svom, Gamow Explorer, THESEUS, and JWST era to understand the nature of diversity among possible progenitors of GRBs at higher redshifts in comparison to those observed in the nearby Universe.

\section{Data availability}
The reduced GTC spectrum and associated tables are available at the CDS via anonymous FTP at [Completed by editor in final version]. All other data are included in the present PDF.

\begin{acknowledgements}
SJ acknowledges the support of the Korea Basic Science Research Program through NRF-2018R1D1A1B07048993, IHP from National Research Foundation grants of 2018R1A2A1A05022685 and 2017K1A4A3015188. MASTER is supported in parts by the Lomonosov Moscow State University Development Program, Moscow Union OPTIKA, and RSF grant 16-12-00085. OG and NB are supported by RFBR grant 17-52-80133. AJCT acknowledges support from the Spanish Ministry project PID2023-151905OB-I00 and Junta de Andaluc\'ia grant P20\_010168 and from the Severo Ochoa grant CEX2021-001131-S funded by MCIN/AEI/10.13039/501100011033. AJCT wishes to express his sincere thanks to W. H. Lee, D. Hiriart, and all members of the San Pedro M\'artir Observatory in M\'exico for their assistance and operations of the 0.6m Javier Gorosabel telescope at the BOOTES-5 station. Data were partly collected with the 10.4m Gran Telescopio Canarias (GTC), installed at the Spanish Observatorio del Roque de los Muchachos of the Instituto de Astrofísica de Canarias, on the island of La Palma. EVK is grateful to the Ministry of Science and Higher Education of the Russian Federation for financial support of this work; the AZT-33IK telescope is part of the Core Shared Research Facility "Angara" of ISTP SB RAS. This work is supported by NASA under an Astrophysics Data Analysis Program (ADAP), and an Astrophysics Theory Program (ATP) funded to UNLV. This work made use of data supplied by the UK Swift Science Data Centre at the University of Leicester. We thank the RATIR project team and the staff of the Observatorio Astronómico Nacional on Sierra San Pedro Mártir and acknowledge the contribution of Leonid Georgiev to its development. RATIR is a collaboration between the University of California, the Universidad Nacional Autonóma de México, NASA Goddard Space Flight Centre, and Arizona State University, benefiting from the loan of an H2RG detector and hardware and software support from Teledyne Scientific and Imaging. RATIR, the automation of the Harold L. Johnson Telescope of the Observatorio Astronómico Nacional on Sierra San Pedro Mártir, and the operation of both are funded through NASA grants NNX09AH71G, NNX09AT02G, NNX10AI27G, and NNX12AE66G, CONACyT grants INFR-2009-01-122785 and CB-2008-101958, NAM PAPIIT grant IG100317, and UC MEXUS-CONACyT grant CN 09-283. SRO gratefully acknowledges the support of the Leverhulme Trust Early Career Fellowship. RS-R acknowledges support from ASI (Italian Space Agency) through the Contract n. 2015-046-R.0 and from the European Union Horizon 2020 Programme under the AHEAD project (grant agreement n. 654215). MCG acknowledges financial support from the Spanish Ministry project MCI/AEI/PID2023-149817OB-C31 and the Severo Ochoa grant CEX2021-001131-S funded by MICIU/AEI/10.13039/501100011033. A.A. acknowledges the Yushan Young Fellow Program by the Ministry of Education, Taiwan, for the financial support (MOE-111-YSFMS-0008-001-P1. SBP, AJCT, and SJ acknowledge valuable contributions from Dr Lucas Uhm during the preparation of the early analysis of \grb data and the presentation. MASTER study was conducted under the state assignment of Lomonosov Moscow State University. OG and NB are supported by the RF Ministry of Science and High Education (projects FZZE-2026-0006, FZZE-2024-0005
\end{acknowledgements}

\bibliographystyle{aa}
\bibliography{ref}

\begin{appendix}
\renewcommand{\thefigure}{A\arabic{figure}}
\renewcommand{\thetable}{A\arabic{table}}
\renewcommand{\thetable}{\thesection.\arabic{table}}
\setcounter{table}{0}
\onecolumn
\section{Prompt emission spectral fitting of \fermi observations} \label{sec:pmt_analysis}
To perform the time-integrated and time-resolved spectra analysis of \fermi-GBM data, we have utilised the multi-mission maximum likelihood (\sw[3ML]; \citealt{2015150708343V}) python package. First, we have downloaded the GBM observations of \grb from the burst catalogue page\footnote{https://heasarc.gsfc.nasa.gov/W3Browse/fermi/fermigbrst.html} of the \fermi science support centre (FSSC). We have utilised the \sw[gtburst] package of \sw[fermitool] to extract the \fermi-GBM spectra. We have utilised the brightest three NaI detectors, n3, n4 and n5 and one BGO detector to extract the spectra in the desired intervals. For the NaI detector, we have utilised the energy range 8-900\,keV, where the energy range 33-37\,keV has been rejected to remove the iodine K-edge. For the BGO detector, the full 200\,keV to 40\,MeV energy range was selected. First, we extracted the time-integrated spectra in a time interval of 0-35\,s. From the observed late X-ray flares, we know that this burst has a long central engine activity; therefore, we have also generated spectra in the time interval of 100-400\,s. However, we have not found any significant GBM detection in this energy range. The fitting results of the time-integrated spectra are listed in Tab. \ref{tab:pmt_param}.\\

To extract the time-resolved spectra, we performed the time slices using the Bayesian binning method, following the instructions of \cite{2014MNRAS.445.2589B}. In this process, we have the 5 time slices, for which we again extract the spectrum using the \sw[gtburst] package. For the spectral fitting in the \sw[3ML], we have loaded all these spectra using the GBM plugin. For spectral fitting, \sw[3ML] provides several inbuilt empirical and physical spectral models. In our case, we have utilised the \sw[Band] and \sw[Cutoff Power-law (CPL)] empirical functions for the spectral fitting and their combination with the physical \sw[Blackbody (BB)] model. We have also utilised a physical \sw[synchrotron] model for the spectral fitting; for more details, please refer to \cite{2020NatAs...4..174B}. To find the best-fit model, we have utilised the deviance information criterion (DIC, \citealt{Spiegelhalter2002}). Since both the \sw[Band] and the \sw[CPL] models are empirical. Further, the \sw[CPL] model is useful when the high-energy spectral index ($\beta$) is poorly constrained \citep{2015AdAst2015E..22P}. But in our case, a \sw[Band] function is appropriate to describe the spectra. Therefore, we have used the \sw[Band] function to compare with the physical synchrotron model. The model with the lowest DIC value is considered the best-fit model. We have found that the \sw[Band+BB] function best describes the observed time-integrated spectrum. In the time-resolved spectral analysis, the significance of the obtained spectra is low, and it is hard to constrain the best-fit model, but the parameters obtained from the \sw[Band] or \sw[Band+BB] model are consistent within error bars. In some spectral bins, the physical \sw[synchrotron] model is also found to best describe the observed spectra. The parameters obtained from the prompt emission spectral fitting are listed in Tab. \ref{tab:pmt_param}. Further, we have retrieved the \fermi-LAT data in the temporal range 0-50000\,s. We have performed the unbinned likelihood analysis of the LAT data following the methodology described in our earlier studies \citep{2025RMxAC..59..133R}. We did not find LAT detection for this burst and constrained an upper limit of LAT flux $\rm<10^{-9}erg~cm^{-2}~s^{-1}$.\\

\begin{figure*}
\centering
\includegraphics[width=0.45\linewidth]{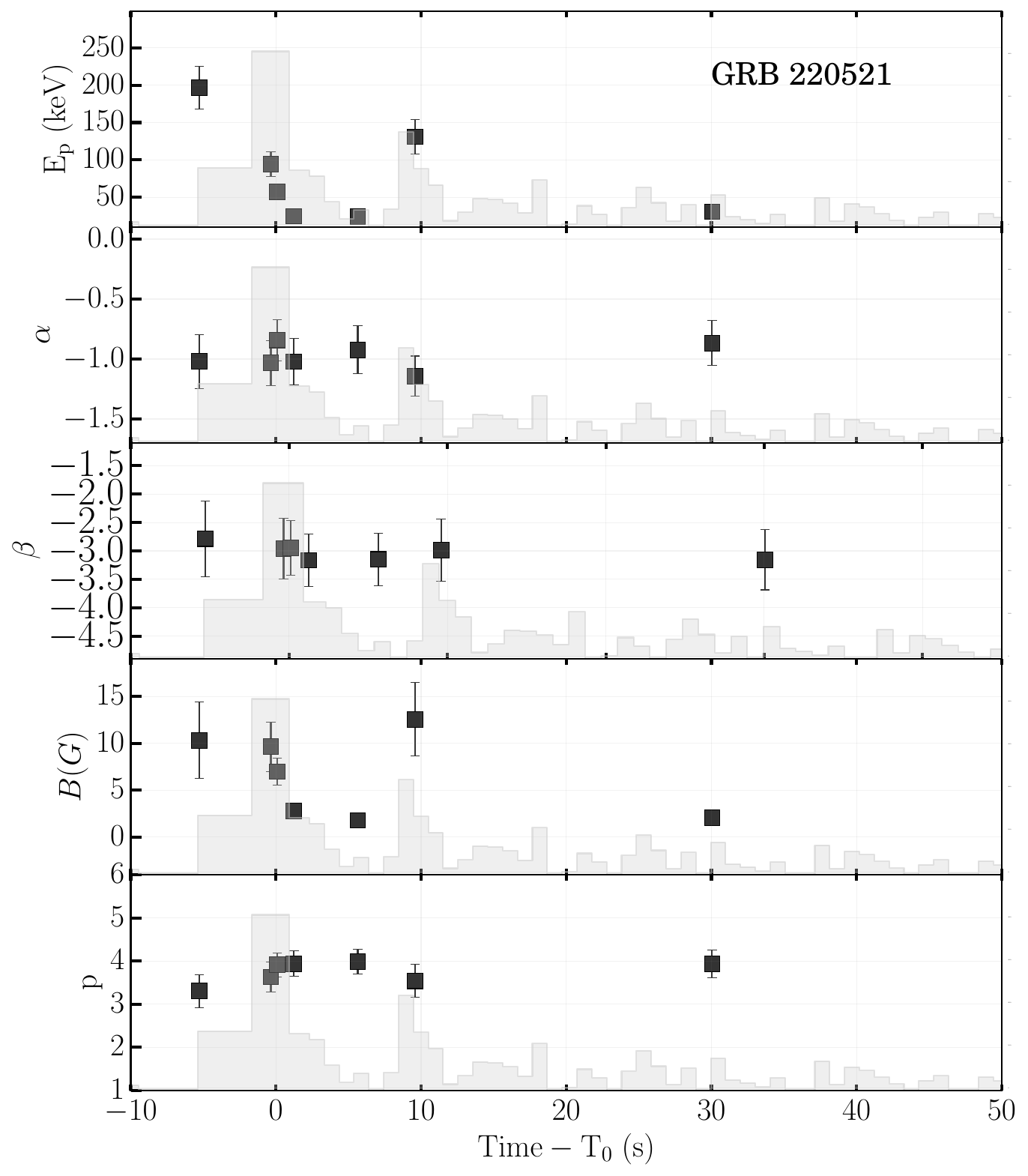}
\includegraphics[width=0.45\linewidth]{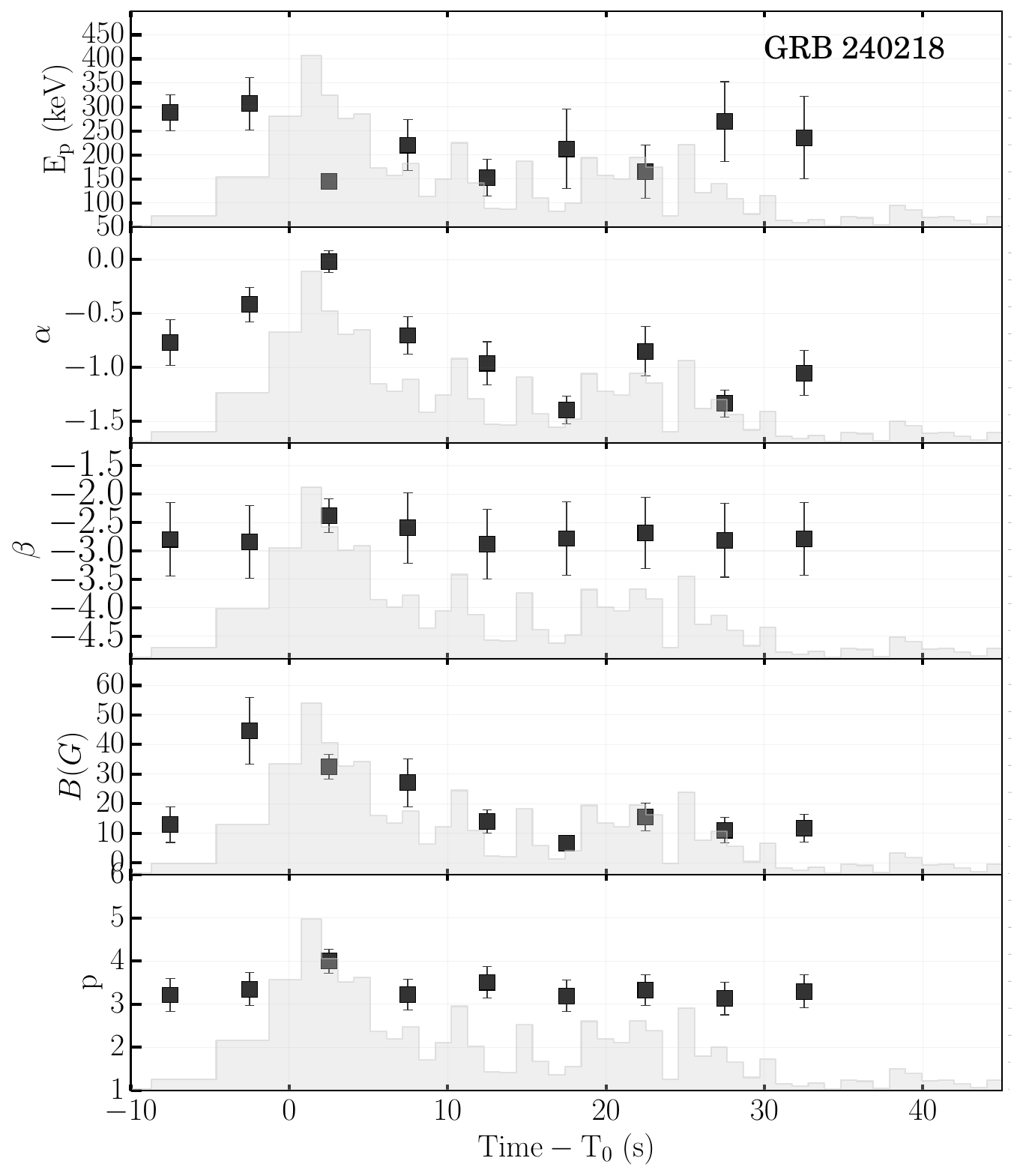}
\caption{Time-resolved spectral parameter evolution of GRB 220521 (left) and GRB 240218 (right), showing the evolution of \Ep, $\alpha$, magnetic field (B), and electron power-law index ($p$) along with the prompt emission light curve overlaid in the background.}
\label{fig:GRB22_24}
\end{figure*}

Further, to check the parameter evolution in other high-$z$ GRBs, we performed the time-resolved spectral analysis for GRB 220521A and GRB 240218A. Most other high-$z$ GRBs either lack \fermi-GBM observations or are too faint for detailed time-resolved spectral studies. The evolution of spectral parameters for these two bursts is shown in Fig. \ref{fig:GRB22_24}. For GRB 220521A, the evolution of spectral displays is not clear, but the obtained spectral parameters appear to follow a hard-to-soft trend. In contrast, for 240218A, the obtained parameters \Ep, $\alpha$ and B are showing a flux tracking evolution. For both the GRBs, $p$ remains relatively stable.

\FloatBarrier
\begin{table*}
\renewcommand{\arraystretch}{1.7}
\centering
\caption{Time integrated and time-resolved prompt emission spectral parameters of \grb obtained from the spectral fitting of \fermi-GBM observations.}
\scriptsize
\begin{tabular}{|c|ccc|cccc|cc|ccc|} \hline
\multicolumn{13}{|c|}{Time Integrated Analysis}\\ \hline
& \multicolumn{3}{c|}{Band} & \multicolumn{4}{c|}{Band+BB} & \multicolumn{2}{c|}{Synchrotron} & \multicolumn{3}{c|}{DIC} \\
time& $\alpha$ & \Ep & $\beta$ & $\alpha$ & Ep & $\beta$ & kT & B & $p$ & DIC$_{Band}$ & DIC$_{Band+BB}$ & DIC$_{syn}$ \\ \hline
12.5 & -0.64$_{-0.22}^{+0.22}$ & 112.08$_{-15.47}^{+14.97}$ & -3.31$_{-0.45}^{+0.44}$ & -0.66$_{-0.22}^{+0.23}$ & 116.17$_{-17.81}^{+15.01}$ & -3.29$_{-0.45}^{+0.45}$& 10.06$_{-5.11}^{+5.12}$ & 14.32$_{-4.01}^{+3.92}$ & 4.18$_{-0.54}^{+0.54}$ & 6329.92 & 6289.75 & 6334.95\\ \hline
\multicolumn{13}{|c|}{Time Resolved Analysis}\\ \hline
& \multicolumn{3}{c|}{Band} & \multicolumn{4}{c|}{Band+BB} & \multicolumn{2}{c|}{Synchrotron} & \multicolumn{3}{c|}{DIC} \\
time& $\alpha$ & \Ep & $\beta$ & $\alpha$ & \Ep & $\beta$ & kT & B & $p$ & DIC$_{Band}$ & DIC$_{Band+BB}$ & DIC$_{syn}$ \\ \hline
-5.33 & -0.96$_{-0.31}^{+0.31}$ & 318.86$_{-142.19}^{+147.82}$ & -2.79$_{-0.66}^{+0.64}$ & $-0.94_{-0.31}^{+0.32}$ & $311.97_{-135.11}^{+144.54}$ & $-2.79_{-0.66}^{+0.64}$ & $9.92_{-4.97}^{+4.83}$ & $28.30_{-18.54}^{+21.63}$ & $3.21_{-0.78}^{+0.78}$ & 4430.61 & 4408.22 & 4345.74\\ 
0.41 & -0.95$_{-0.22}^{+0.21}$ & 232.39$_{-77.52}^{+78.05}$ & -2.84$_{-0.65}^{+0.62}$ & $-0.95_{-0.21}^{+0.22}$ & $236.14_{-78.00}^{+77.86}$ & $-2.84_{-0.63}^{+0.61}$ & $10.00_{-4.99}^{+4.90}$ & $24.76_{-13.16}^{+15.06}$ & $3.41_{-0.73}^{+0.73}$ & 2321.17 & 2302.65 & 2272.22\\ 
4.08 & -1.03$_{-0.43}^{+0.43}$ & 148.08$_{-80.76}^{+85.43}$ & -2.94$_{-0.61}^{+0.59}$ & $-0.97_{-0.42}^{+0.43}$ & $237.35_{-152.59}^{+177.52}$ & $-2.87_{-0.64}^{+0.63}$ & $11.86_{-3.85}^{+3.49}$ & $11.03_{-6.99}^{+6.25}$ & $3.50_{-0.76}^{+0.76}$ & 3565.95 & 3560.45 & 3511.46\\ 
9.68 & -0.69$_{-0.23}^{+0.23}$ & 136.94$_{-25.18}^{+23.69}$ & -3.01$_{-0.56}^{+0.52}$ & $-0.72_{-0.23}^{+0.23}$ & $148.15_{-33.30}^{+26.73}$ & $-3.02_{-0.55}^{+0.52}$ & $10.49_{-5.22}^{+5.06}$ & $17.84_{-6.05}^{+5.77}$ & $3.88_{-0.61}^{+0.63}$ & 3811.35 & 3769.86 & 3809.32\\ 
31.34 & -0.77$_{-0.443}^{+0.43}$ & 192.34$_{-132.02}^{+140.12}$ & -2.83$_{-0.65}^{+0.64}$ & $-0.92_{-0.44}^{+0.43}$ & $287.56_{-172.93}^{+175.75}$ & $-2.79_{-0.64}^{+0.63}$ & $9.66_{-4.74}^{+4.72}$ & $9.41_{-7.93}^{+6.79}$ & $3.27_{-0.75}^{+0.78}$ & 6454.4 & 6409.82 & 6452.24\\ \hline
& \multicolumn{3}{c|}{CPL} & \multicolumn{4}{c|}{CPL+BB} & \multicolumn{2}{c|}{} & \multicolumn{3}{c|}{} \\
time & $\alpha$ & Ep & & $\alpha$ & \Ep & & kT & & & DIC$_{CPL}$ & DIC$_{CPL+BB}$ & \\ \hline
-5.33 & $-1.07_{-0.23}^{0.23}$ & $345.41_{-154.47}^{153.95}$ & & $-0.97_{-0.25}^{0.26}$ & $324.24_{-153.16}^{155.87}$ & & $10.68_{-6.64}^{6.49}$ & & & 4427.48 & 4395.18 & \\
0.41 & $-1.04_{-0.16}^{0.16}$ & $275.21_{-109.45}^{111.41}$ & & $-0.99_{-0.17}^{0.17}$ & $260.97_{-103.78}^{105.54}$ & & $10.84_{-6.58}^{6.45}$ & & & 2317.77 & 2293.26 & \\
4.08 & $-1.32_{-0.31}^{0.31}$ & $234.18_{-145.14}^{157.31}$ & & $-1.12_{-0.42}^{0.43}$ & $281.11_{-182.47}^{188.46}$ & & $13.24_{-3.54}^{3.73}$ & & & 3573.72 & 3554.75 & \\
9.68 & $-0.85_{-0.19}^{0.19}$ & $144.81_{-49.03}^{47.06}$ & & $-0.84_{-0.22}^{0.22}$ & $159.35_{-64.49}^{58.69}$ & & $11.82_{-7.27}^{6.83}$ & & & 3791.81 & 3742.49 & \\
31.34 & $-1.32_{-0.41}^{0.41}$ & $320.31_{-182.12}^{180.49}$ & &$-1.11_{-0.44}^{0.44}$ & $313.31_{-181.51}^{180.28}$ & & $10.05_{-6.18}^{6.04}$ & & & 6452.72 & 6422.53 &  \\ \hline
\end{tabular}
\label{tab:pmt_param}
\end{table*}

\section{Prompt emission spectral fitting of \swift observations}
Using the \swift-BAT observation, we have fitted the prompt emission spectrum of \grb using a \sw[power-law], a \sw[Cutoff power-law], and a \sw[Band] function \citep{1993ApJ...413..281B}. Since the BAT has a soft energy range of 15-150\,keV, generally a \sw[power-law] function is found to best describe the data. The photon index obtained from the fitting of the \sw[Power-law] function to the Joint BAT and XRT spectrum yields a photon index $\Gamma_{BAT} = 1.46 \pm 0.05$. We have also fitted a \sw[Cutoff power-law] and a \sw[Band] function to the observed Joint BAT and XRT spectrum. The \sw[Cutoff power-law] function results in the photon Index $\Gamma_{BAT} = 1.44 \pm 0.05$ and the cutoff energy $E_{c} = 530 \pm 200$. Obviously, the observed $E_{c}$ is not in the BAT energy range. Similarly, fitting of the \sw[Band] function results in the low energy decay index $\alpha = 0.59 \pm 0.17$, peak energy \Ep = $88 \pm 12$, and the high energy decay index $\beta$ = $2.12 \pm 0.22$. We have used the deviance information criteria (DIC) to find the best-fit model. The \sw[Power-law] model has the lowest DIC value of 838, which is less than the value obtained from fitting the \sw[Band] function DIC = 918 and the \sw[Cutoff power-law] function DIC = 851. We have also tried to fit the physical \sw[synchrotron] model to the observations, but the model parameters could not be constrained due to the soft energy range of \swift-BAT.

\begin{figure*}
\centering
\includegraphics[width=0.9\columnwidth]{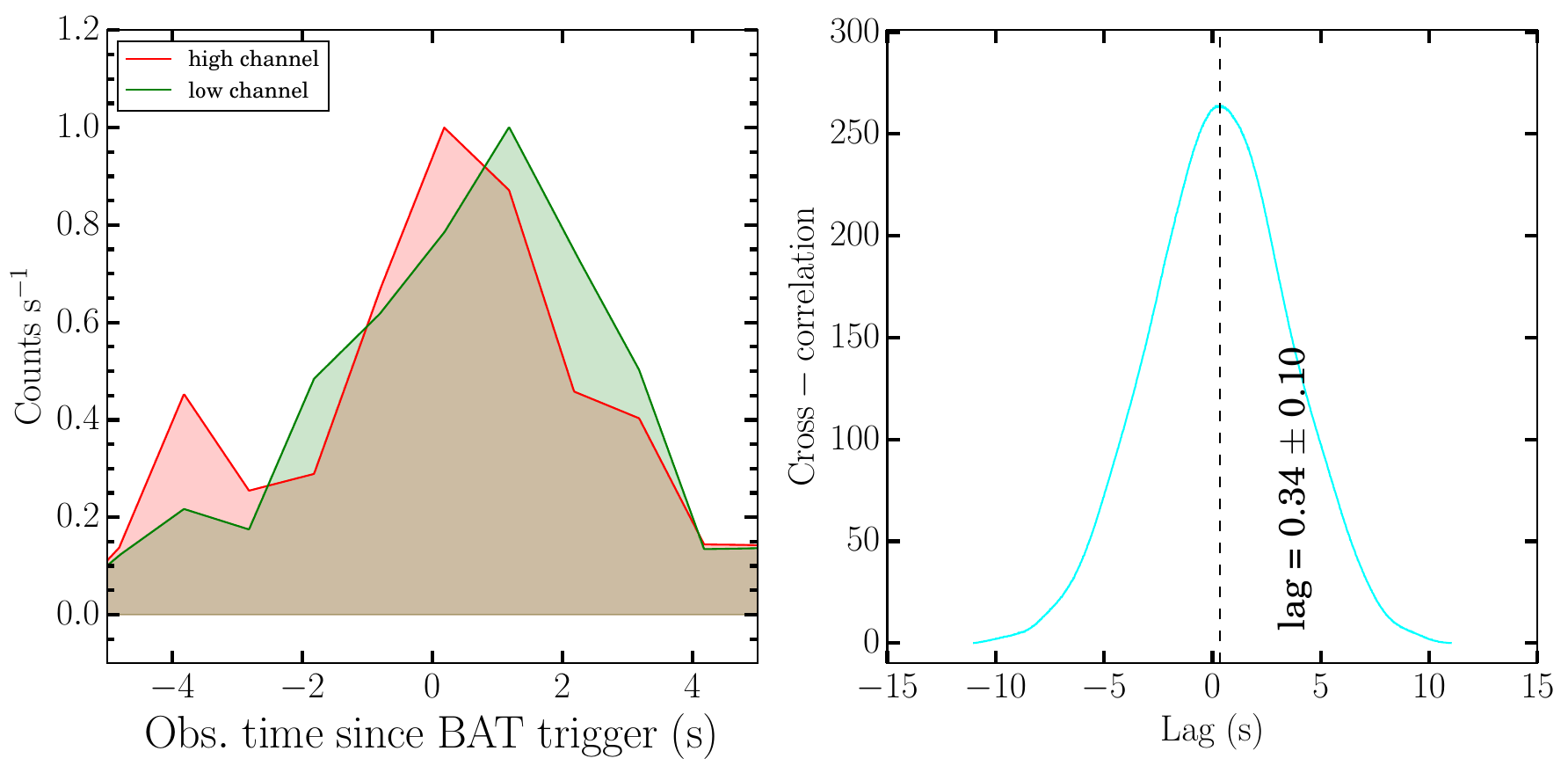}
\caption{Time lag calculated between two energy channels \((50-100)\)\,keV and \((15-25)\)\,keV of prompt emission \swift-BAT light curve of \grb.}
\label{fig:CC_lag}
\end{figure*}

\begin{figure*}
\centering
\includegraphics[width=0.48\columnwidth]{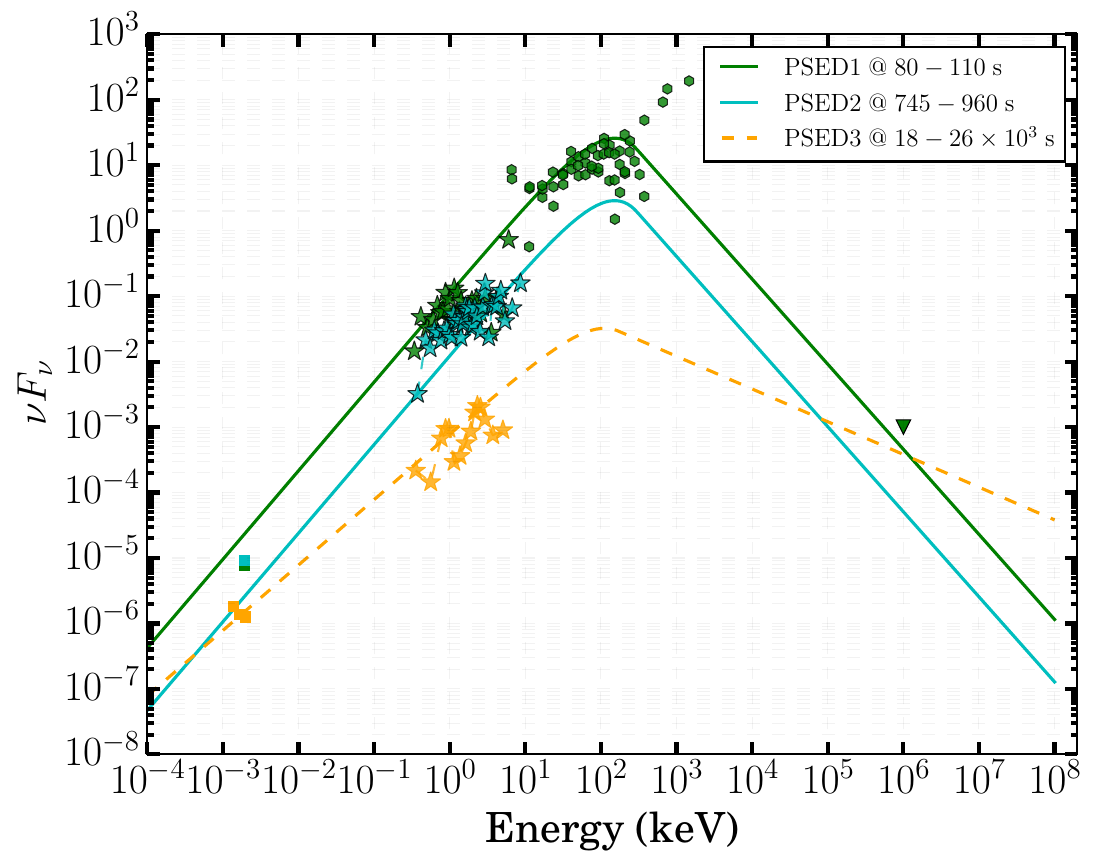}
\caption{Spectral energy distribution created using GBM data during prompt emission and the late-time optical, X-ray, and GeV observation at three epochs (PSED1, PSED2, PSED3), as shown in the legend. The band function given in the solid line was obtained by fitting the prompt emission spectra using the GBM observation. The dashed curve represents the Band function, using typical parameters, to satisfy the data point corresponding to the last epoch (PSED3).}
\label{fig:PSED}
\end{figure*}

\begin{table*}
\centering
\caption{The fitting parameters of the simultaneous flares observed in optical, X-ray (0.3-10 keV; XRT) and $\gamma$-ray (15-150 keV; \swift-BAT) bands.}
\begin{tabular}{|c|cc|cc|ccc|} \hline
{Flares} & \multicolumn{2}{c|}{{Optical}} & \multicolumn{2}{c|}{{XRT}} & \multicolumn{3}{c|}{{BAT}} \\ \hline
& FWHM (s) & Peak position (s) & FWHM (s) & Peak position (s) & FWHM (s) & Peak position (s) & Significance ($\sigma$) \\
\hline
0 & ND & ND & 148 $\pm$ 24 & 161 $\pm$ 7 & ND & ND & ND \\
1 & 328 $\pm$ 112 & 535 $\pm$ 27 & 121 $\pm$ 11 & 374 $\pm$ 5 & 66 $\pm$ 39 & 302 $\pm$ 21 & 2.7 \\
2 & 547 $\pm$ 75 & 1115 $\pm$ 20 & 118 $\pm$ 11 & 824 $\pm$ 5 & 66 $\pm$ 16 & 637 $\pm$ 22 & 1.9 \\
3 & 762 $\pm$ 154 & 1650 $\pm$ 50 & ND & ND & ND & ND & ND \\
\hline
\end{tabular}
\label{tab:G_fit}
\end{table*}

\begin{figure*}
\centering
\includegraphics[width=0.9\columnwidth]{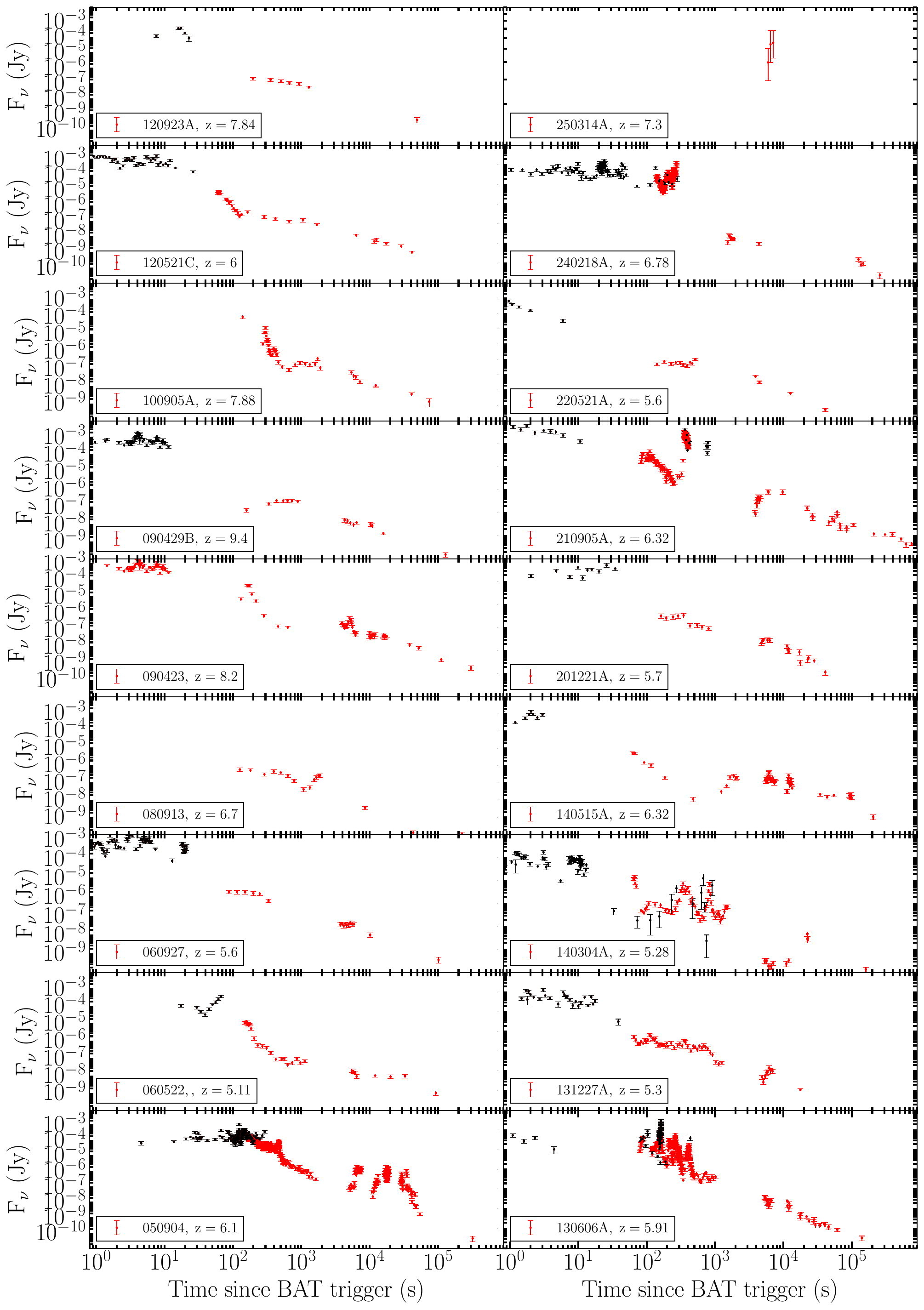}
\caption{The comparison between the prompt (BAT @ 10\,keV) and afterglow (XRT @ 10\,keV) flux density light curve taken from \swift burst analyser \citep{eva07, eva09} webpage of high redshift GRBs with \(z > 5\).}
\label{fig:bat_xrt_lcs}
\end{figure*}

\begin{table*}
\centering
\caption{The log and photometry of optical observations of \grb.}
\scriptsize
\begin{tabular}{ccccc||ccccc} \hline
T$_{0}$-T$_{mid}$ (s) & Telescope & Exp. & Filter & Magnitude (mag) & T$_{0}$-T$_{mid}$ (s) & Telescope & Exp. & Filter & Magnitude (mag) \\ \hline
82.9 & MASTER-II & 10 & P\_EAST & 16.35$\pm$0.19 & 5740.9 & Mondy(AZT-33IK) & 120$\times$10 & \textit{R} & 21.07$\pm$0.14\\
89.3 & MASTER-II & 10 & P\_WEST & 15.98$\pm$0.32 & 89705 & Mondy(AZT-33IK) & 120$\times$60 & \textit{I} & $>$19.90\\
172.2 & MASTER-II & 30 & P\_EAST & 17.18$\pm$0.17 & 929.26 & Nanshan & 60 & \textit{R} & 19.01$\pm$0.14\\
172.9 & MASTER-II & 30 & P\_WEST & 17.64$\pm$0.30 & 1315.4 & Nanshan & 60 & \textit{R} & 19.40$\pm$0.20\\
294.2 & MASTER-II & 50 & P\_EAST & 18.00$^{+0.52}_{-0.35}$ & 1464.5 & Nanshan & 60 & \textit{R} & 19.76$\pm$0.27\\
295.9 & MASTER-II & 50 & P\_WEST & 18.41$^{+0.81}_{-0.46}$ & 1614.7 & Nanshan & 60 & \textit{R} & 19.59$\pm$0.22\\
413.7 & MASTER-II & 70 & P\_EAST & 18.12$\pm$0.29 & 1764.2 & Nanshan & 60 & \textit{R} & 19.52$\pm$0.21\\
415.5 & MASTER-II & 70 & P\_WEST & 19.09$\pm$0.76 & 2555.4 & Nanshan & 150 & \textit{R} & 20.22$\pm$0.26\\
556 & MASTER-II & 100 & P\_EAST & 18.29$\pm$0.31 & 3220.2 & Nanshan & 150 & \textit{R} & 20.31$\pm$0.30\\
556.4 & MASTER-II & 100 & P\_WEST & 18.39$^{+0.52}_{-0.35}$ & 24581.5 & GTC & 30 & \textit{r} & 20.27$\pm$0.03\\
719.8 & MASTER-II & 130 & P\_EAST & 18.85$^{+0.92}_{-0.49}$ & 25333.4 & GTC & 60 & \textit{r} & 20.26$\pm$0.04\\
720.5 & MASTER-II & 130 & P\_WEST & 18.90$^{+1.52}_{-0.61}$ & 24055.1 & GTC & 10 & \textit{i} & 18.79$\pm$0.04\\
937.8 & MASTER-II & 170 & P\_EAST & 18.63$\pm$0.31 & 24204.7 & GTC & 10 & \textit{i} & 18.74$\pm$0.03\\
963.5 & MASTER-II & 170 & P\_WEST & 19.15$^{+1.60}_{-0.62}$ & 24341.4 & GTC & 20 & \textit{i} & 18.77$\pm$0.02\\
1193.3 & MASTER-II & 180 & P\_EAST & 18.63$\pm$0.29 & 24499 & GTC & 30 & \textit{i} & 18.70$\pm$0.03\\
1195 & MASTER-II & 180 & P\_WEST & 18.72$^{+0.54}_{-0.36}$ & 25446.9 & GTC & 60 & \textit{i} & 18.77$\pm$0.03\\
454.9 & Khureltogot(ORI-40) & 60 & None & 17.78$\pm$0.28 & 25603.2 & GTC & 60 & \textit{i} & 18.78$\pm$0.02\\
519.9 & Khureltogot(ORI-40) & 60 & None & 17.97$\pm$0.22 & 24669.4 & GTC & 30 & \textit{z} & 17.46$\pm$0.05\\
584.9 & Khureltogot(ORI-40) & 60 & None & 17.94$\pm$0.20 & 25208.9 & GTC & 60 & \textit{z} & 17.49$\pm$0.06\\
649.9 & Khureltogot(ORI-40) & 60 & None & 18.25$\pm$0.25 & 49254.9 & RATIR & 1280 & \textit{r} & 21.63$\pm$0.13\\
744.9 & Khureltogot(ORI-40) & 60$\times$2 & None & 18.57$\pm$0.26 & 135824.9 & RATIR & 1280 & \textit{r} & 22.62$\pm$0.00\\
874.9 & Khureltogot(ORI-40) & 60$\times$2 & None & 18.99$\pm$0.24 & 49254.9 & RATIR & 1280 & \textit{i} & 20.28$\pm$0.07\\
974.9 & Khureltogot(ORI-40) & 60 & None & 18.05$\pm$0.24 & 50884.5 & RATIR & 1280 & \textit{i} & 20.29$\pm$0.08\\
1038.9 & Khureltogot(ORI-40) & 60 & None & 18.49$\pm$0.28 & 135824.9 & RATIR & 1280 & \textit{i} & 22.28$\pm$0.00\\
1115.9 & Khureltogot(ORI-40) & 60 & None & 18.18$\pm$0.28 & 48969 & RATIR & 536.9 & \textit{z} & 18.90$\pm$0.06\\
1180.9 & Khureltogot(ORI-40) & 60 & None & 18.18$\pm$0.20 & 50596 & RATIR & 536.9 & \textit{z} & 18.94$\pm$0.07\\
1395.9 & Khureltogot(ORI-40) & 60$\times$6 & None & 18.92$\pm$0.27 & 135540 & RATIR & 536.9 & \textit{z} & 20.35$\pm$0.20\\
2018.9 & Mondy(AZT-33IK) & 60$\times$4 & \textit{R} & 19.97$\pm$0.08 & 105.5 & BOOTES-4 & 0.5 & C & $>$15.8\\
2258.9 & Mondy(AZT-33IK) & 60$\times$4 & \textit{R} & 20.10$\pm$0.09 & 3370.3 & BOOTES-4 & 1467.5 & C & $>$18.3\\
2528.9 & Mondy(AZT-33IK) & 60$\times$5 & \textit{R} & 20.04$\pm$0.07 & 3108.2 & BOOTES-4 & 820 & \textit{g} & $>$19.8\\
3038.9 & Mondy(AZT-33IK) & 12$\times$5 & \textit{R} & 20.19$\pm$0.08 & 3387.9 & BOOTES-4 & 780 & \textit{r} & $>$19.1\\
3639.9 & Mondy(AZT-33IK) & 120$\times$5 & \textit{R} & 20.68$\pm$0.13 & 4034.2 & BOOTES-4 & 600 & \textit{i} & $>$19.2\\
4539.9 & Mondy(AZT-33IK) & 120$\times$10 & \textit{R} & 20.53$\pm$0.10 & 4186.4 & BOOTES-4 & 1140 & \textit{z} & $>$19.0\\ \hline
\end{tabular}
\tablefoot{All magnitudes are in the Vega system and are not corrected for Galactic extinction.}
\label{tab:optical_obs}
\end{table*}

\end{appendix}

\end{document}